\documentclass[aps,prl,twocolumn,amsmath,amssymb,superscriptaddress,10pt]{revtex4-1}
\usepackage{graphicx}
\usepackage{ulem}
\usepackage{soul}
\usepackage{txfonts}
\usepackage[colorlinks, linkcolor=blue]{hyperref}
\usepackage{verbatim}
\usepackage{esint}
\usepackage{amsmath}
\usepackage{mathtools}
\usepackage{float}
\restylefloat{table}

\def \ve {\varepsilon}
\def \beq {\begin{eqnarray}}
\def \eeq {\end{eqnarray}}
\def \tn {\textnormal}

\begin{document}

\title{Strange metal in magic-angle graphene with near Planckian dissipation}
\author{Yuan Cao}
\thanks{{\bf These authors contributed equally to this project.}}
\author{Debanjan Chowdhury}
\thanks{{\bf These authors contributed equally to this project.}}
\author{Daniel Rodan-Legrain}
\author{Oriol Rubies-Bigord\`a}
\affiliation{Department of Physics, Massachusetts Institute of Technology, Cambridge Massachusetts
02139, USA.}
\author{Kenji Watanabe}
\author{Takashi Taniguchi}
\affiliation{National Institute for Materials Science, Namiki 1-1, Tsukuba, Ibaraki 305-0044, Japan.}
\author{T. Senthil}
\thanks{{\bf To whom correspondence should be addressed; E-mail: senthil@mit.edu, pjarillo@mit.edu.}}
\author{Pablo Jarillo-Herrero}
\thanks{{\bf To whom correspondence should be addressed; E-mail: senthil@mit.edu, pjarillo@mit.edu.}}
\affiliation{Department of Physics, Massachusetts Institute of Technology, Cambridge Massachusetts
02139, USA.}

\date{\today \\
\vspace{.1in}}

\begin{abstract}
{\bf Recent experiments on magic angle twisted bilayer graphene have discovered correlated insulating behavior and superconductivity at a fractional filling of an isolated narrow band. In this paper we show that magic angle bilayer graphene exhibits another hallmark of strongly correlated systems --- a broad regime of $T-$linear resistivity above a small, density dependent, crossover temperature--- for a range of fillings near the correlated insulator. We also extract a transport ``scattering rate", which satisfies a near Planckian form that is universally related to the ratio of $(k_BT/\hbar)$. Our results establish magic angle bilayer graphene as a highly tunable platform to investigate strange metal behavior, which could shed light on this mysterious ubiquitous phase of correlated matter. }    
\end{abstract}

\maketitle

A panoply of strongly correlated materials have metallic parent states that display properties at odds with the expectations in a conventional Fermi liquid and are marked by the absence of coherent quasiparticle excitations. Some well known families of materials, such as the ruthenates~\cite{hussey,kapitulnik}, cobaltates~\cite{Taillefer1,Ong1} and a subset of the iron-based superconductors~\cite{Matsuda14}, show non-Fermi liquid (NFL) behavior over a broad intermediate range of temperatures, with a crossover to a conventional Fermi liquid below an emergent low-energy scale, $T_{\tn{coh}}$, at which coherent electronic quasiparticles emerge as well-defined excitations. Even more striking examples of NFL behavior are observed in the  hole-doped cuprates  \cite{Takagi,boebinger} and certain quantum critical heavy-fermion compounds  \cite{rmpqcp} where the incoherent features appear to survive down to the lowest measurable temperatures, i.e. $T_{\tn{coh}}\rightarrow0$, when superconductivity is suppressed externally. In the incoherent regime, all of these materials in spite of being microscopically distinct have a resistivity, $\rho(T)\sim T$, and exhibit a number of anomalous features \cite{zxs,Marel,johnson,rmpqcp} clearly indicating the absence of sharp electronic quasiparticles. In strongly interacting non-quasiparticle systems, it has been conjectured \cite{Zaanen04} that transport ``scattering rates" ($\Gamma$) satisfy a universal `Planckian' bound  $\Gamma\lesssim O(k_BT/\hbar)$ at a temperature $T$.  It is however worth noting that in a NFL there is in general no clear definition for $\Gamma$; the scattering rates defined through different measurements, such as dc and optical conductivity, need not be identical \cite{DC2018}. One of the most surprising aspects of incoherent transport in these systems, in spite of these subtleties, is that $\Gamma$ extracted from the dc resistivity (through a procedure specified in Ref. \cite{Bruin13}) appears to satisfy a universal form $\Gamma=C k_BT/\hbar$ with $C$ a number of order 1 \cite{Bruin13,Taillefer18}. 

\begin{figure*}
\begin{center}
\includegraphics[scale=0.55]{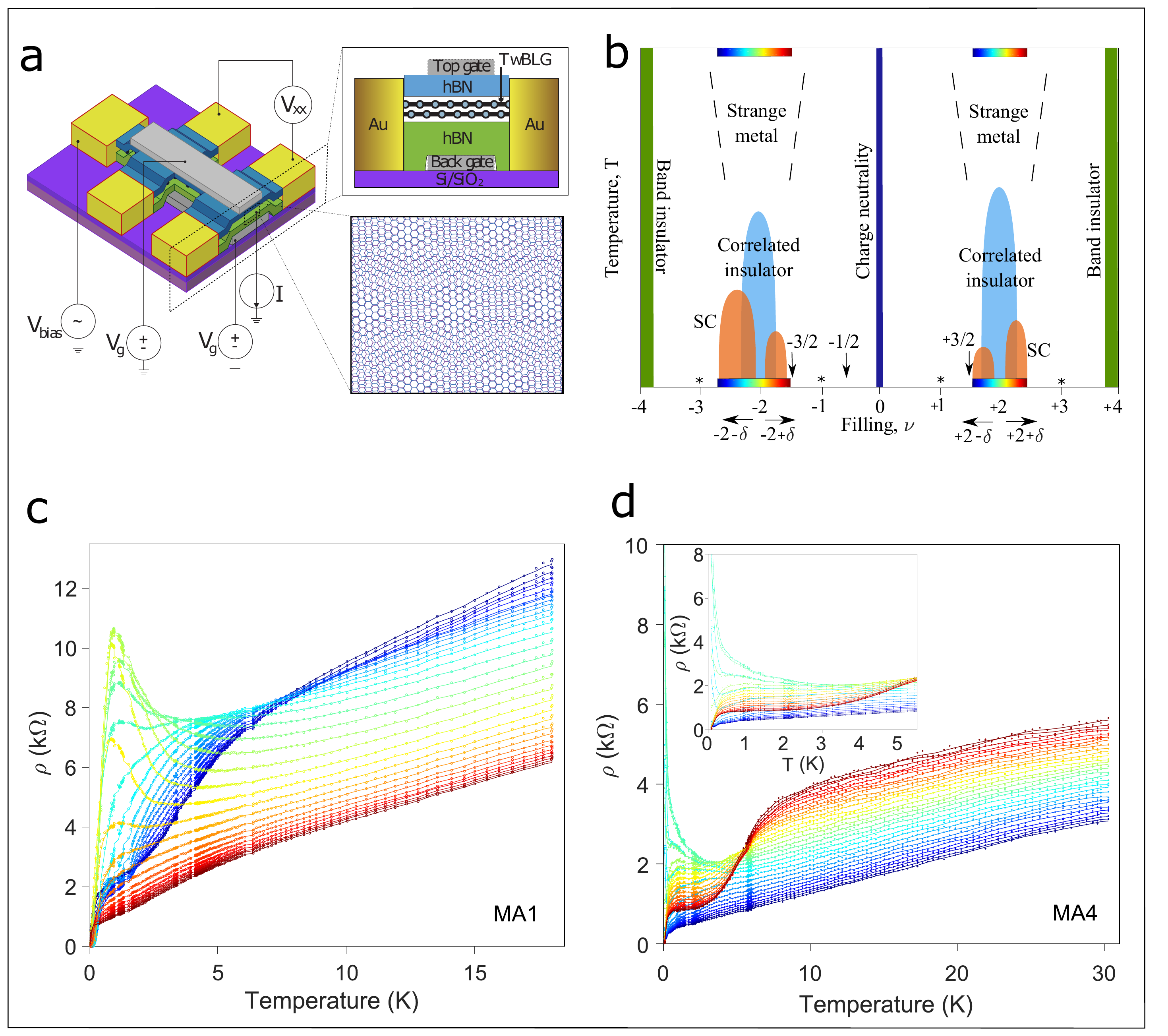}
\end{center}
\caption{{\bf Transport in MABLG. a,} Device schematic and four-probe transport measurement configuration (left). Top-right: side-view of the device structure, consisting of bilayers of graphene twisted by $\theta$ relative to each other, sandwiched by hBN on the top and bottom. The carrier density is tuned using the
metal gates at the top/bottom. Bottom-right: schematic MABLG moir$\acute{\tn{e}}$ pattern. {\bf b,} A schematic phase diagram for MABLG as a function of temperature and filling, $\nu$. The strange metal behavior with $T-$linear resistivity is primarily restricted to fillings near $\nu=\pm2$. Orange regions denote superconductivity. Light blue regions denote the correlated insulator region. The color bars indicate the approximate filling ranges investigated and shown in Fig. 1c,d. {\bf c,} Resistivity ($\rho$) as a function of temperature for device MA1 ($\theta=1.16^0$) for gate-induced densities from $-1.96\times 10^{12}$ cm$^{-2}$ (blue) to $-1.22\times 10^{12}$ cm$^{-2}$ (red); see horizontal colorbar near $\nu=-2$ in Fig.\ref{rho}b. {\bf d,} $\rho(T)$ for device MA4 ($\theta=1.16^0$) for gate-induced densities ranging from $1.23\times 10^{12}$ cm$^{-2}$ (blue) to $1.89\times 10^{12}$ cm$^{-2}$ (red); see horizontal colorbar near $\nu=+2$ in Fig.\ref{rho}b. The inset shows the traces for the same device at low temperatures. The solid smooth lines have been obtained by using a  gaussian-weighted filter.}
\label{rho}
\end{figure*}

Recent experiments \cite{Cao2018a,Cao2018b} have reported the discovery of a correlation driven insulator at fractional fillings (with respect to a fully filled isolated band) in magic-angle bilayer graphene (MABLG). In MABLG, the relative rotation between two sheets of graphene generates a moir$\acute{\tn{e}}$ pattern  (Fig. \ref{rho}a) with a periodicity that is much larger than the underlying interatomic distances in graphene. The theoretically estimated electronic bandwidth, $W$, is strongly renormalized near these small magic-angles \cite{Castro,suarez,macdonald11}; then the strength of the typical Coulomb interactions, $U$, becomes at least comparable to (if not greater than) the bandwidth, $U\gtrsim W$. Investigating the properties of MABLG as a function of temperature and carrier density with unprecedented tunability in a controlled setup can lead to new insights into the nature of electronic transport in other low-dimensional strongly correlated metallic systems. 

For our experiments, we have fabricated multiple high-quality encapsulated MABLG devices (see Fig.~\ref{rho}a) using the `tear and stack' technique \cite{Tutuc,Cao2016}. As reported earlier \cite{Cao2016,Cao2018a}, we obtain band-insulators with a large gap near
$n\approx\pm n_s$, where $n$ is the carrier density tuned externally by applying a gate voltage and $n_s$ corresponds to four electrons per moir$\acute{\tn{e}}$ unit cell. On the other hand, correlation driven insulators with much smaller gap-scales \cite{Cao2018a,dean} appear near $n \approx \pm n_s/2$; we denote these fillings as $\nu=\pm 2$ from now on (the fully filled band corresponds to $\nu = +4$).  Doping away from these correlated insulators by an additional amount, $\delta$, with holes ($\nu=\pm 2 - \delta$) and electrons ($\nu=\pm 2 +\delta$) leads to superconductivity (SC)~\cite{Cao2018b,dean}. The superconducting transition temperature ($T_c$) measured relative to the Fermi-energy ($\ve_F$), as inferred from low temperature quantum oscillations  measurements~\cite{Cao2018b}, is high, with the largest value of $(T_c/\ve_F)\sim 0.07 - 0.08$, indicating strong coupling superconductivity~\cite{Cao2018b}. A schematic $\nu - T$ phase-diagram for MABLG is shown in Fig.\ref{rho}b.

In this work, we investigate the transport phenomenology of the metallic states in MABLG as a function of increasing temperature over a large range of externally tuned fillings.  We have analyzed the temperature dependence of the longitudinal DC resistivity, $\rho(T)$, for a number of devices (MA1 - MA6) over a wide range of fillings, the results of which appear to be qualitatively similar across devices. In order to highlight the universal aspects of the behavior, both within and across different samples, we show the temperature dependent traces of $\rho(T)$ for a range of different $\nu$ in two devices MA1 and MA4 in Fig.\ref{rho}c and Fig.\ref{rho}d respectively. The range of $\nu$ chosen for this purpose is shown as a color-bar in Fig.\ref{rho}b; see caption for details. As is evident from these traces, at low temperatures the resistivity exhibits highly non-monotonic features as a function of $\nu$, reflecting the complicated and distinct nature of the ground states \cite{shoulder} (see also inset of Fig. \ref{rho}d). On the other hand, at higher temperatures, the traces for both devices look qualitatively similar and show a clear monotonic trend as a function of the gate-tuned filling in spite of the underlying low-temperature differences. As $\nu$ is tuned externally  starting from near the superlattice density, the number of carriers increases resulting in an enhanced conductivity. This simple picture is to be taken seriously only at high temperatures; at low temperatures the effective number of carriers changes non-monotonically across $\nu=-2$ \cite{Cao2018b}. We identify the characteristic temperature scale associated with the crossover between these distinct regimes, which depends on $\nu$ itself, as $T_{\tn{coh}}(\nu)$. For a wide range of fillings in the vicinity of $\nu=\pm 2$, we find that $\rho(T)\sim AT$ for $T\gtrsim T_{\tn{coh}}(\nu)$, where $A$ denotes the slope of the resistivity in units of ohms/K ($\Omega/$K). In a given device, $A$ has a relatively weak dependence on $\nu$, but there is a substantial variation in the value of A for fillings $\nu=\pm 2$ for a given device, as well as in between devices (which have invariably slightly different twist angles, $\theta$). 

\begin{figure*}
\begin{center}
\includegraphics[scale=0.7]{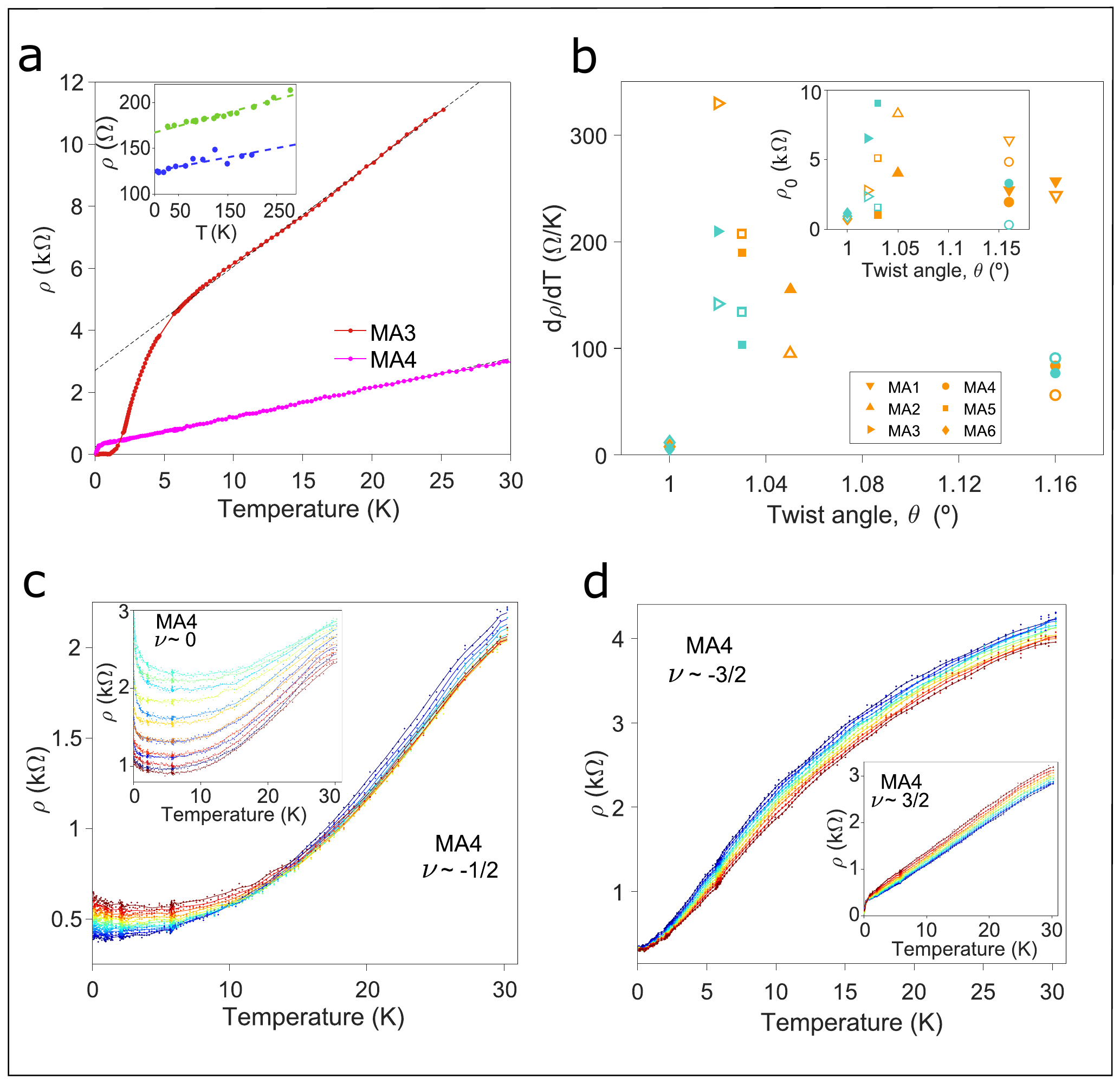}
\end{center}
\caption{{\bf $T-$linear resistivity near $\nu=\pm2$ in MABLG. a,} Resistivity ($\rho$) as a function of temperature for device MA3 ($\theta=1.02^0$) and MA4 ($\theta=1.16^0$) at a gate-tuned density of $-1.46\times10^{12}$ cm$^{-2}$ ($\nu= -2-\delta$) and $1.19\times 10^{12}$ cm$^{-2}$ ($\nu=2-\delta$), respectively. The inset shows $\rho(T)$ in ohms for MLG on SiO$_2$ (green) and hBN (blue), respectively (data from~\cite{monolayerT,mlghbn}). {\bf b,} The slopes $A=d\rho/dT$ obtained at the fillings near $\nu=\pm2\pm\delta$ with optimal superconducting $T_c$ for six different devices (MA1 - MA6) as a function of respective twist-angles. Orange (cyan) markers denote fillings with $\nu>0$ ($\nu<0$), while solid (empty) symbols denote deviations, $\delta>0$ ($\delta<0$), away from the correlated insulators at $\nu=\pm2$. The inset shows the extrapolated resistivity, $\rho_0$, for the same devices.  {\bf c,} $\rho(T)$ for device MA4 near $\nu=-1/2$ for gate-induced densities $-0.51$ to $-0.29\times10^{12}$ cm$^{-2}$. The data look similar near $\nu=+1/2$. The inset shows $\rho(T)$ for the same device near $\nu=0$ and densities between $\pm0. 13\times10^{12}$ cm$^{-2}$. {\bf d,} $\rho(T)$ for device MA4 near $\nu=-3/2$ for gate-induced densities $-1.28$ to $-1.08\times 10^{12}$ cm$^{-2}$. The inset shows $\rho(T)$ for device MA4 near $\nu=+3/2$ over a symmetric range of densities.}
\label{quant}
\end{figure*}

Fig. \ref{quant}a shows $\rho(T)$ for two separate MABLG devices, MA3 and MA4, at optimal hole doping, i.e. at fillings where the measured $T_c$ is highest on the hole doped side of the correlated insulators at $\nu=-2-\delta$ (MA3) and $\nu=2-\delta$ (MA4). From a linear fit to the regime $T>T_{\tn{coh}}$, we obtain $A\approx335 ~\Omega/$K for MA3 and $A\approx 95~\Omega/$K for MA4. It is interesting to note that a naive estimate for the slope of the resistivity, assuming  $\rho(T)\sim AT$ and accounting for the spin and valley degeneracies at high temperatures, would lead to a value of $A\sim (h/4e^2W)$, up to $O(1)$ prefactors \cite{supp}. For an estimated bandwidth of order $W\sim 10$ meV, this gives $A\approx 60~\Omega/$K which is close to the observed values. We now compare and contrast the above results for MABLG with monolayer graphene (MLG), a chemically similar, but weakly correlated metallic system, that also displays $T-$linear resistivity extending over a much wider range of temperatures. In the inset of Fig. \ref{quant}a, we show two representative examples of $\rho(T)$ for MLG on a SiO$_2$ \cite{monolayerT} and hBN \cite{mlghbn} substrate, respectively. The resistivity in MLG devices is approximately two orders of magnitude smaller than MABLG, making it a significantly more conducting metal compared to MABLG. Similarly, the slope of the resistivity for MLG on both substrates is $A\approx 0.1 ~\Omega/$K and is nearly doping independent for a range of different densities near charge neutrality. The slope in MLG is thus approximately three to four orders of magnitude smaller than the corresponding slope in MABLG. Since there is some variation in the value of the slopes across different samples, we have studied the dependence of $A$ on the twist angle, $\theta$, as shown in Fig.~\ref{quant}b. The slopes are always evaluated for optimal doping, i.e. the filling at which $T_c$ is the highest. Orange (cyan) markers denote fillings with $\nu>0$ ($\nu<0$), while solid (empty) symbols denote deviations, $\delta>0$ ($\delta<0$), away from the correlated insulators at $\nu=\pm2$. As it is clear from the figure, the slopes at a given $\theta$ depend on the sign of $\nu$. Experimentally, it is unclear at what value of $\theta$ the magic-angle condition is satisfied precisely. However, we find the highest superconducting $T_c$ around $\theta\sim 1.03^0$, which might possibly indicate closest proximity to the magic-angle. While we do not have enough data-points to obtain a clear trend, the typical values of the slopes are large near $\theta=1.02^0-1.03^0$ (MA3, MA5, respectively) and $\theta=1.16^0$ (MA1, MA4). Fillings $\nu>0$ also tend to have higher slopes than $\nu<0$. We can also obtain the intercept, $\rho_0=\rho(T\rightarrow0)$, from an extrapolation of the data from high temperatures which gives us a rough estimate of the residual resistivity due to elastic scattering. The results are shown in the inset of Fig. \ref{quant}b.   

One of the most interesting aspects of the longitudinal transport data in MABLG is that the $T-$linear resistivity is primarily confined to fillings near the correlated insulators at $\nu=\pm2\pm\delta$ with $\delta\lesssim1/2$. We note however that we do observe a number of correlated phenomena~\cite{dean}, and some unusual transport properties including $T-$linear resistivity, also near $\nu=\pm1,\pm3$ (marked by asterisks in Fig.\ref{rho}b) in some of our devices, results that will be reported elsewhere. Fig. \ref{quant}c shows the traces of $\rho(T)$ in the vicinity of $\nu=-1/2$ (see corresponding arrow on the schematic phase-diagram in Fig. \ref{rho}b) for device MA4 ; the traces at $\nu=+1/2$ look qualitatively similar. The inset of Fig. \ref{quant}c shows $\rho(T)$ near charge neutrality at $\nu=0$. It is clear that $\rho(T)$ exhibits qualitatively distinct behavior near these fillings, without any clear indication of a broad regime of $T-$linear resistivity. In Fig.\ref{quant}d and its inset we plot $\rho(T)$ near $\nu=-3/2$ and $\nu=+3/2$ for device MA4 (fillings marked by arrows in Fig.\ref{rho}b), respectively. While for fillings near $\nu=-3/2$ there is no apparent $T-$linear resistivity, there is some indication of such behavior near $\nu=3/2$. 

One of the most celebrated examples of strange metal behavior is observed in the hole-doped cuprates near optimal doping. It shows a number of remarkable features, including $\rho(T)\sim BT$ over hundreds of kelvin without any apparent signs of crossovers \cite{Takagi,boebinger,Bruin13,Taillefer18}, anomalous power-law dependent optical conductivities \cite{Marel} and incoherent spectral functions in the presence of a sharp Fermi surface \cite{zxs}. Focusing on the transport properties, the typical value of the in-plane sheet resistivity in LSCO (which depends on sample quality) at $T=300~$K is approximately $\rho \sim 0.2 - 0.3~$ m$\Omega~$cm \cite{Takagi,boebinger}, which when normalized by the appropriate interplane distance ($d\sim 6.4$ angstroms) leads to a value of the resistivity, $\rho\sim 3.1 \times 10^3 - 4.7\times 10^3~\Omega$. These values are roughly comparable to the measured values of the resistivity in MABLG. 
On the other hand, the value of the slope in LSCO is approximately $B\sim 1.0 - 2.0 ~\mu\Omega~$cm/K \cite{Takagi,boebinger}, which when normalized by $d$ leads to $B\sim 15.6 - 31.2~\Omega/$K. The slope is thus slightly smaller than the typical values in MABLG.

\begin{figure*}
\begin{center}
\includegraphics[scale=0.82]{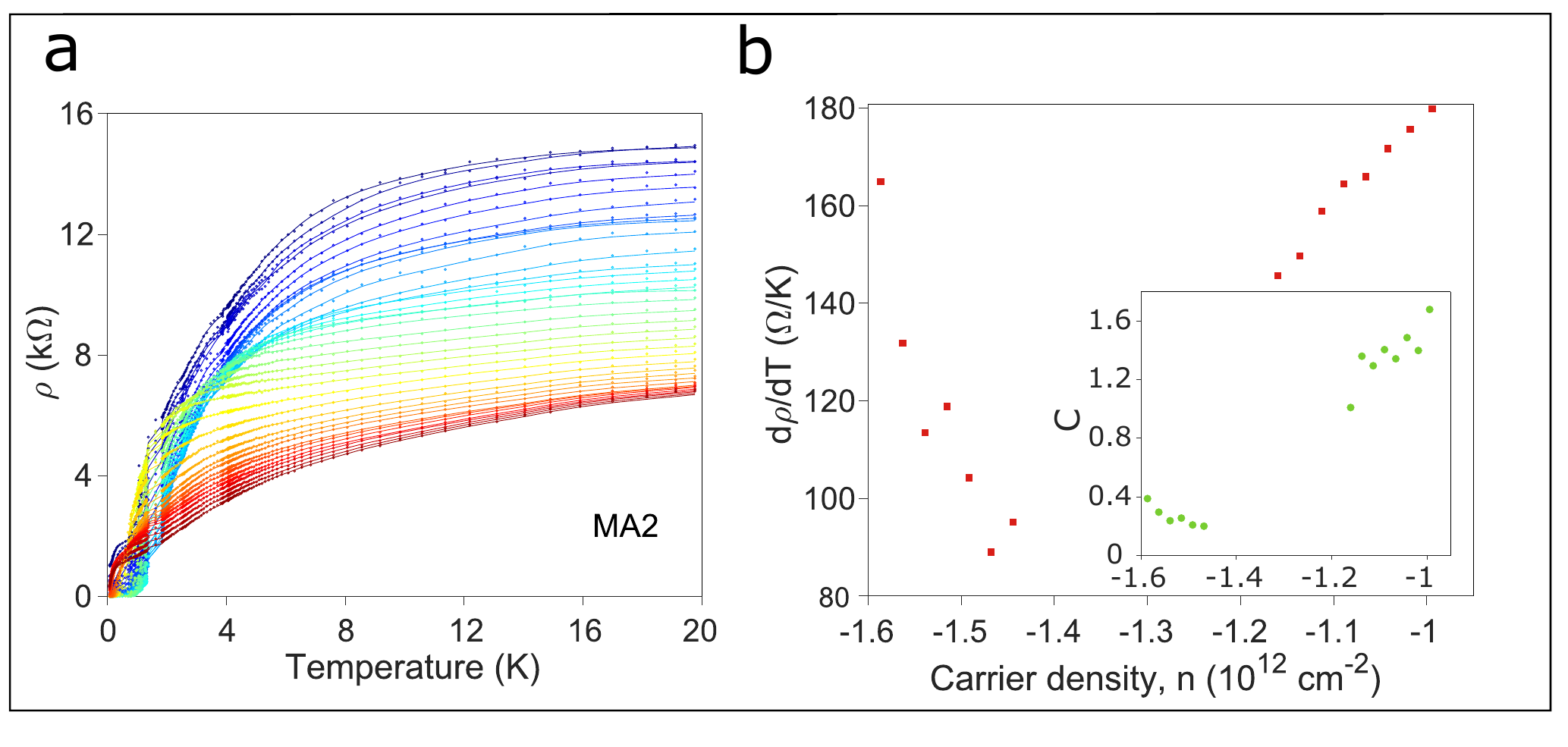}
\end{center}
\caption{{\bf Planckian scattering in magic-angle bilayer graphene. a,} Resistivity as a function of temperature for device MA2 ($\theta=1.05^0$) near $\nu=-2$ for gate-tuned densities $-1.85 \times 10^{12}$ cm$^{-2}$ (blue) to $-1.00\times10^{12}$ cm$^{-2}$ (red). The correlated insulator at $\nu=-2$ is approximately located  near $-1.3\times10^{12}$ cm$^{-2}$. The range of fillings is shown using the horizontal colorbar near $\nu=-2$ in Fig.\ref{rho}a. The solid smooth lines have been obtained by using a  gaussian-weighted filter. {\bf b,} Slope of resistivity as a function of filling above $8~$K for the same device evaluated using a linear fit. The inset shows the coefficient $C$ of the scattering rate $\Gamma = C k_BT/\hbar$ for the same fillings as in Fig. \ref{ana}b, respectively.}
\label{ana}
\end{figure*}

Thus far we have compared the typical (non-universal) values of the resistivities and their slopes across different MABLG devices, as well as with other materials such as MLG and optimally doped cuprates. We now focus on a more detailed presentation and analysis of device MA2. We begin by showing the traces of $\rho(T)$ for MA2 in Fig.\ref{ana}a for a range of $\nu=-2\pm\delta$, roughly corresponding to the color-bar in Fig.\ref{rho}b. As in the other devices, we observe that the resistivity at high temperatures decreases monotonically as $\nu$ increases towards charge neutrality, presumably due to the increase in the number of carriers measured relative to the superlattice filling. We have evaluated the slope $A=d\rho(T)/dT$ at high temperatures for the same range of $\nu$ considered in Fig.\ref{ana}a above. Fig.\ref{ana}b shows the values of $A$ as a function of the gate-induced carrier density for fillings near $\nu=-2$. The slope here is obtained directly from a linear least-squares fit to the resistivity  \cite{supp} for $T>8~$K. It is interesting to note that the slope varies non-monotonically with $\nu$.   On the other hand, it is possible that a transport ``scattering rate" ($\Gamma$) exhibits more universal character. In order to define and obtain $\Gamma$ from just transport data within the same device, we shall now adopt the procedure used in Ref.\cite{Bruin13}. In the regime of $T-$linear resistivity, we  write the scattering rate, $\Gamma = C k_BT/\hbar$, and  {\it define} the numerical prefactor $C$ as a function of $\nu$ for device MA2 by assuming that the resistivity can be well described by a Drude-like expression \cite{Bruin13,Taillefer18}, such that
\beq
C = \frac{\hbar}{k_B}\frac{e^2~n_c(0)}{m^*(0)} A,
\label{planckian}
\eeq
where $n_c(0),~m^*(0)$ are the measured density and effective mass values at low temperatures, $T\rightarrow0$.  This expression gives an operational definition of $C$, and hence of the scattering rate $\Gamma$. In device MA2, we have also been able to extract the {\it actual} low-$T$ carrier density ($n_c(0)$) and effective mass ($m^*(0)$) from   Shubnikov-de Haas (SdH) quantum oscillations measurements \cite{Cao2018b}. Note that $n_c(0)$, as inferred from SdH measurements, is not simply proportional to $\nu$ everywhere in the phase-diagram. Instead for $\nu=-2-\delta$,  $n_c(0)$ corresponds to the gate-induced density relative to the correlated insulator at $\nu=-2$\cite{Cao2018b}. On the other hand, for $\nu=-2+\delta$,  $n_c(0)$ corresponds to the gate-induced density relative to charge-neutrality at $\nu=0$ \cite{Cao2018b}, in agreement with low-temperature Hall measurements.  From the SdH measurements, we can estimate the Fermi-temperature, $T_F = 2\pi\hbar^2 n_c(0)/(k_Bgm^*(0))$, where $g$ is the degeneracy factor. Interestingly, $\rho(T)$ does not exhibit any characteristic changes or crossovers as the temperature approaches $T_F$ for fillings near $\nu=-2-\delta$, where $T_F\lesssim 30$ K \cite{supp}. In all other examples of strange metals in solid-state systems, the range over which $T-$linear resistivity is observed is significantly below $T_F$. MABLG is thus unique in that the $T-$linear resistivity appears to persist to the electron degeneracy temperature, as inferred from the low temperature measurements.

It is remarkable that $C\sim O(1)$ number for all $\nu$ studied near $\nu=-2$ (see inset of Fig.~\ref{ana}). Moreover, $C$ is weakly dependent on the filling for $\nu=-2-\delta$ and varies from $C\sim 0.2 - 0.4$. On the other hand, for  $\nu=-2+\delta$, the coefficient increases monotonically with increasing $\delta$, with $C\sim 1.0 - 1.6$. It is interesting to note that the superconducting $T_c$ is lower for the fillings where $\Gamma$ is larger \cite{supp}. A similar analysis can not be carried out asymptotically close to the insulating filling $\nu=-2$ because of the absence of low-temperature SdH measurements of $m^*$ and $n_c$. However, since the slope in the metallic regime at high temperatures evolves smoothly as a function of $\delta$ near $\nu=-2$ and the value of $n_c/m^*$ is almost independent of the filling, we may naively use the extrapolated value of this ratio for $\delta\rightarrow0$ to infer that the high-temperature metal at $\nu=-2$ has a similar form of scattering rate.

A similar procedure when used to determine the scattering rates for other strongly correlated metals \cite{Bruin13,Taillefer18}, including ones that lead to unconventional superconductivity at low temperatures,  also leads to $C\sim O(1)$ number. A major advantage of our analysis for MABLG is that all of our measurements are carried out on the same device (MA2) to extract $C$, as opposed to using data from multiple devices at different dopings.  For various members of the cuprate family, the coefficient $C\sim 0 .7 - 1.2$, both on the hole and electron-doped side of the phase-diagram \cite{Bruin13,Taillefer18}. It is worth mentioning that $C\sim O(1)$ number even for metals like copper, aluminum and palladium above their respective Debye temperatures, where they exhibit a $T-$linear resistivity due to electron-phonon scattering  \cite{Bruin13}. In MLG, the $T-$linear resistivity is also likely due to electron-phonon scattering above a density dependent temperature \cite{SDSRMP}. We can calculate the scattering rates for MLG on SiO$_2$ in the regime where they display $T-$linear resistivity. Let us express the resistivity (in units of $h/e^2$) as
\beq
\rho(T) = \frac{h}{e^2}\frac{\hbar\Gamma}{2\ve_F(V)},
\eeq
where $\Gamma$ is as defined earlier and $\ve_F(V)$ is controlled by an external gate-voltage, $V$. This leads to an approximate estimate of $C\sim 0.009 - 0.022$ for the entire range of densities studied in Ref.~\cite{monolayerT}, which is a few orders of magnitude smaller than the value in MABLG. In Table \ref{C}, we show a few representative examples of systems where the scattering rate has been calculated using the above procedure.

We now critically examine some possible mechanisms that may be responsible for the unusual transport phenomenology in MABLG. Recently, it has been suggested that electron-phonon scattering may be sufficient to explain the origin of $T-$linear resistivity in twisted BLG near the magic-angle \cite{SDS18}. This theory is applicable primarily in the vicinity of fillings $\nu=0$ and the coefficient of the high temperature $\rho\sim T$ regime is then controlled by the strongly renormalized band-structure near charge-neutrality (Dirac-points). However, in our devices, a clear evidence for $\rho\sim T$ and Planckian scattering rates can only be seen near the commensurate fillings of $\nu=\pm 2\pm\delta$ with $\delta\lesssim1/2$, but not in the vicinity of charge neutrality $|\nu|\lesssim1/2$. Moreover, $T_{\tn{coh}}(\nu)$ can be as low as $0.5~$K (e.g. for device MA4 near $\nu=-2$ in Fig.\ref{quant}a), which is an order of magnitude lower than the predicted crossover temperature \cite{SDS18}. The maximum value of the slope of the $T-$linear resistivity, which has a weak dependence on $\nu$ in a given device, is observed near the angle ($\theta\sim1.03^0$) as well as at an angle of $\theta\sim 1.16^0$ (Fig.\ref{quant}b). However, ref.\cite{SDS18} predicts a very strong angle dependence of the slope, with the largest value near the magic-angle, and this dependence is tied to the strong renormalization of the Dirac velocity. Although we do not observe this trend, such a discrepancy could arise as a result of long-wavelength disorder and slow variations in the twist-angle across our devices. In addition, since the $T-$linear resistivity persists to scales $T\sim T_F$ without any signs of a crossover near $\nu=-2-\delta$, while the electrons are no longer in the degenerate limit, electron-phonon scattering in the metallic state $(T>T_c)$ is likely insufficient to explain the origin of these unusual results. In principle, $T-$linear resistivity can arise as a result of scattering off any electronic collective mode fluctuations (instead of phonons) for $T_M\lesssim T$, where $T_M$ is a characteristic scale associated with the respective collective mode, but a crossover across $T\sim T_F$ is nevertheless expected. It is unclear at the moment if such electronic collective modes are present near $\nu=\pm 2$. 

Under conditions of local equilibrium, the conductivity is given by $\sigma = \chi D_c$, where $\chi=\partial n/\partial\mu$ is the electronic compressibility and $D_c\sim v^2/\Gamma$ is the charge diffusivity ($v\equiv$ a characteristic velocity). Thus far, we have inherently assumed that the temperature dependence of the resistivity arises as a result of the temperature dependence of $\Gamma\sim k_BT/\hbar$.  An alternative scenario that could, in principle, lead to $\rho\sim T$ at high temperatures arises solely from the temperature dependence of the compressibility, $\chi\sim1/T$, in a system with a bounded kinetic energy. However, in some of our devices the $\rho\sim A T$ behavior persists from high temperatures down to temperatures as low as $0.5~$K without any change in the slope, $A$ (e.g. device MA4 in Fig.~\ref{quant}a). Moreover, as emphasized earlier, $\rho$ does not show any crossovers as the temperature is swept through the estimated $\ve_F$ for fillings near $\nu=-2-\delta$.  Thus while it is unlikely that the above mechanism is responsible for the observed temperature dependence of the resistivity, a systematic future study of the charge compressibility in MABLG as a function of increasing temperature is highly desirable. Instead of relying on the dc transport measurements to extract $\Gamma$, it will be interesting to measure and analyze the extent to which the optical scattering rates in MABLG satisfy a universal form in future experiments. Interestingly, recent measurements of diffusivity and compressibility in a cold-atoms based experiment \cite{bakr} find $\rho\sim T$ from temperatures higher than the bandwidth down to lower temperatures without any signs of a crossover.

In light of these results, any successful theory has to account for the following universal aspects of the phenomenology: (i) a $T-$linear resistivity with values $O(h/e^2)$ in the vicinity of commensurate fillings,  $\nu=\pm2$, with near Planckian ($C\sim O(1)$) ``scattering rates", (ii) a weak dependence of the slope of the resistivity on $\nu$, (iii) relative insensitivity of the resistivity to $T_F$ and by extension, to the underlying details of the Fermi-surface at low temperatures, and (iv) the presence of a small $T_{\tn{coh}}$ above which the transport is unconventional. Since $T_{\tn{coh}}$ can be as low as $0.5~$K in some of our devices, it will be interesting to see if future experiments find evidence of $T-$linear resistivity down to even lower temperatures (i.e. $T_{\tn{coh}}\rightarrow0$). If so, it is possible that the NFL behavior in MABLG is controlled by a $T=0$ quantum critical point or quantum critical phase. On the other hand, if $T_{\tn{coh}}$ is non-zero, it is likely that the metallic regime of MABLG for fillings near $\nu=\pm2$ realizes an intermediate-scale NFL. Assuming that the temperature remains smaller than the typical interaction strengths but large compared to $T_{\tn{coh}}$, the state is neither a classical liquid (or gas) nor a degenerate quantum liquid. The system can then be best described as a ``semi-quantum" liquid, with no coherent quasiparticle excitations and no sharply defined Fermi surface. Describing such a regime in a theoretically controlled limit is challenging but recent progress has been made in the study of some models \cite{Hartnoll18,DC2018} which find evidence of such incoherent behavior. Guided by these studies and by the present experiments we will further develop microscopic theories of transport in such a regime elsewhere. 

\begin{table}[H]
\begin{tabular}{||l| l | c | c|r ||}
  \hline \hline
 &  & Slope ($A$) &   &  \\
 & Material & 3D: $\mu\Omega$ cm / K &  $C$& Refs.\\
 & & 2D: $\Omega$ / K & & \\
  \hline \hline
  & CeCoIn$_5$ & $1.6$ & $1$ & \cite{Bruin13}\\
  & CeRu$_2$Si$_2$ & $0.91$ & $1.1$ & \cite{Bruin13}\\
3D  & (TMTSF)$_2$PF$_6$ & $0.38$ & $0.9$ & \cite{Bruin13}\\
 & ($11.8$ kbar)  & & & \\
  & UPt$_3$ & $1.1$ & $1.1$ & \cite{Bruin13} \\
  & Cu ($T>100$ K) & $7\times 10^{-3}$ & $1.0$ & \cite{Bruin13}\\
  & Au ($T>100$ K) & $8.4\times 10^{-3}$ & $0.96$ & \cite{Bruin13}\\
  \hline
  & Bi2212 ($p=0.22$) & $8.0$ & $1.1$ & \cite{Taillefer18} \\
 & LSCO ($p=0.26$) & $8.2$ & $0.9$ & \cite{Taillefer18}\\
(Quasi-)2D & PCCO ($x=0.17$) & $1.7$ & $1.0$ & \cite{Taillefer18}\\
  & MLG on SiO$_2$ & $0.1$ & $0.01 - 0.02$ & \cite{monolayerT}, This work \\
  & MABLG & $100 - 300$ & $0.2 - 1.6$ & This work   \\
  \hline \hline 
\end{tabular}
\caption{Materials exhibiting Planckian scattering rates ($\Gamma$), where $\Gamma = Ck_BT/\hbar$. }
\label{C}
\end{table}

\section*{Acknowledgements}
This work has been primarily supported by the National Science Foundation (DMR-1809802), the Center for Integrated Quantum Materials under NSF grant DMR-1231319, and the Gordon and Betty Moore Foundation's EPiQS Initiative through Grant GBMF4541 to P.J.-H. for device fabrication, transport measurements, and data analysis. This work was performed in part at the Harvard University Center for Nanoscale Systems (CNS), a member of the National Nanotechnology Coordinated Infrastructure Network (NNCI), which is supported by the National Science Foundation under NSF ECCS award no. 1541959. D.C. is supported by a postdoctoral fellowship from the Gordon and Betty Moore Foundation, under the EPiQS initiative, grant GBMF4303, at MIT. D.R.-L acknowledges support from Obra Social ``la Caixa" Fellowship. O.R.-B acknowledges support from Fundaci$\acute{\tn{o}}$ Privada Cellex. T.S. is supported by a US Department of Energy grant DE-SC0008739, and in part by a Simons Investigator award from the Simons Foundation. Growth of BN crystals was supported by the Elemental Strategy Initiative conducted by the MEXT, Japan and JSPS KAKENHI grant numbers JP15K21722 and JP25106006.

\bibliographystyle{apsrev4-1_custom}
\bibliography{strangemetal}

\begin{thebibliography}{34}%
\makeatletter
\providecommand \@ifxundefined [1]{%
 \@ifx{#1\undefined}
}%
\providecommand \@ifnum [1]{%
 \ifnum #1\expandafter \@firstoftwo
 \else \expandafter \@secondoftwo
 \fi
}%
\providecommand \@ifx [1]{%
 \ifx #1\expandafter \@firstoftwo
 \else \expandafter \@secondoftwo
 \fi
}%
\providecommand \natexlab [1]{#1}%
\providecommand \enquote  [1]{``#1''}%
\providecommand \bibnamefont  [1]{#1}%
\providecommand \bibfnamefont [1]{#1}%
\providecommand \citenamefont [1]{#1}%
\providecommand \href@noop [0]{\@secondoftwo}%
\providecommand \href [0]{\begingroup \@sanitize@url \@href}%
\providecommand \@href[1]{\@@startlink{#1}\@@href}%
\providecommand \@@href[1]{\endgroup#1\@@endlink}%
\providecommand \@sanitize@url [0]{\catcode `\\12\catcode `\$12\catcode
  `\&12\catcode `\#12\catcode `\^12\catcode `\_12\catcode `\%12\relax}%
\providecommand \@@startlink[1]{}%
\providecommand \@@endlink[0]{}%
\providecommand \url  [0]{\begingroup\@sanitize@url \@url }%
\providecommand \@url [1]{\endgroup\@href {#1}{\urlprefix }}%
\providecommand \urlprefix  [0]{URL }%
\providecommand \Eprint [0]{\href }%
\providecommand \doibase [0]{http://dx.doi.org/}%
\providecommand \selectlanguage [0]{\@gobble}%
\providecommand \bibinfo  [0]{\@secondoftwo}%
\providecommand \bibfield  [0]{\@secondoftwo}%
\providecommand \translation [1]{[#1]}%
\providecommand \BibitemOpen [0]{}%
\providecommand \bibitemStop [0]{}%
\providecommand \bibitemNoStop [0]{.\EOS\space}%
\providecommand \EOS [0]{\spacefactor3000\relax}%
\providecommand \BibitemShut  [1]{\csname bibitem#1\endcsname}%
\let\auto@bib@innerbib\@empty
\bibitem [{\citenamefont {Hussey}\ \emph {et~al.}(1998)\citenamefont {Hussey},
  \citenamefont {Mackenzie}, \citenamefont {Cooper}, \citenamefont {Maeno},
  \citenamefont {Nishizaki},\ and\ \citenamefont {Fujita}}]{hussey}%
  \BibitemOpen
  \bibfield  {author} {\bibinfo {author} {\bibfnamefont {N.~E.}\ \bibnamefont
  {Hussey}}, \bibinfo {author} {\bibfnamefont {A.~P.}\ \bibnamefont
  {Mackenzie}}, \bibinfo {author} {\bibfnamefont {J.~R.}\ \bibnamefont
  {Cooper}}, \bibinfo {author} {\bibfnamefont {Y.}~\bibnamefont {Maeno}},
  \bibinfo {author} {\bibfnamefont {S.}~\bibnamefont {Nishizaki}}, \ and\
  \bibinfo {author} {\bibfnamefont {T.}~\bibnamefont {Fujita}},\ }\bibfield
  {title} {\enquote {\bibinfo {title} {Normal-state magnetoresistance of
  ${\mathrm{sr}}_{2}{\mathrm{ruo}}_{4}$},}\ }\href {\doibase
  10.1103/PhysRevB.57.5505} {\bibfield  {journal} {\bibinfo  {journal} {Phys.
  Rev. B}\ }\textbf {\bibinfo {volume} {57}},\ \bibinfo {pages} {5505}
  (\bibinfo {year} {1998})}\BibitemShut {NoStop}%
\bibitem [{\citenamefont {Klein}\ \emph {et~al.}(1996)\citenamefont {Klein},
  \citenamefont {Dodge}, \citenamefont {Ahn}, \citenamefont {Snyder},
  \citenamefont {Geballe}, \citenamefont {Beasley},\ and\ \citenamefont
  {Kapitulnik}}]{kapitulnik}%
  \BibitemOpen
  \bibfield  {author} {\bibinfo {author} {\bibfnamefont {L.}~\bibnamefont
  {Klein}}, \bibinfo {author} {\bibfnamefont {J.~S.}\ \bibnamefont {Dodge}},
  \bibinfo {author} {\bibfnamefont {C.~H.}\ \bibnamefont {Ahn}}, \bibinfo
  {author} {\bibfnamefont {G.~J.}\ \bibnamefont {Snyder}}, \bibinfo {author}
  {\bibfnamefont {T.~H.}\ \bibnamefont {Geballe}}, \bibinfo {author}
  {\bibfnamefont {M.~R.}\ \bibnamefont {Beasley}}, \ and\ \bibinfo {author}
  {\bibfnamefont {A.}~\bibnamefont {Kapitulnik}},\ }\bibfield  {title}
  {\enquote {\bibinfo {title} {Anomalous spin scattering effects in the badly
  metallic itinerant ferromagnet ${\mathrm{srruo}}_{3}$},}\ }\href {\doibase
  10.1103/PhysRevLett.77.2774} {\bibfield  {journal} {\bibinfo  {journal}
  {Phys. Rev. Lett.}\ }\textbf {\bibinfo {volume} {77}},\ \bibinfo {pages}
  {2774} (\bibinfo {year} {1996})}\BibitemShut {NoStop}%
\bibitem [{\citenamefont {Li}\ \emph {et~al.}(2004)\citenamefont {Li},
  \citenamefont {Taillefer}, \citenamefont {Hawthorn}, \citenamefont {Tanatar},
  \citenamefont {Paglione}, \citenamefont {Sutherland}, \citenamefont {Hill},
  \citenamefont {Wang},\ and\ \citenamefont {Chen}}]{Taillefer1}%
  \BibitemOpen
  \bibfield  {author} {\bibinfo {author} {\bibfnamefont {S.~Y.}\ \bibnamefont
  {Li}}, \bibinfo {author} {\bibfnamefont {L.}~\bibnamefont {Taillefer}},
  \bibinfo {author} {\bibfnamefont {D.~G.}\ \bibnamefont {Hawthorn}}, \bibinfo
  {author} {\bibfnamefont {M.~A.}\ \bibnamefont {Tanatar}}, \bibinfo {author}
  {\bibfnamefont {J.}~\bibnamefont {Paglione}}, \bibinfo {author}
  {\bibfnamefont {M.}~\bibnamefont {Sutherland}}, \bibinfo {author}
  {\bibfnamefont {R.~W.}\ \bibnamefont {Hill}}, \bibinfo {author}
  {\bibfnamefont {C.~H.}\ \bibnamefont {Wang}}, \ and\ \bibinfo {author}
  {\bibfnamefont {X.~H.}\ \bibnamefont {Chen}},\ }\bibfield  {title} {\enquote
  {\bibinfo {title} {Giant electron-electron scattering in the fermi-liquid
  state of
  ${\mathrm{n}\mathrm{a}}_{0.7}\mathrm{C}\mathrm{o}{\mathrm{o}}_{2}$},}\ }\href
  {\doibase 10.1103/PhysRevLett.93.056401} {\bibfield  {journal} {\bibinfo
  {journal} {Phys. Rev. Lett.}\ }\textbf {\bibinfo {volume} {93}},\ \bibinfo
  {pages} {056401} (\bibinfo {year} {2004})}\BibitemShut {NoStop}%
\bibitem [{\citenamefont {Wang}\ \emph {et~al.}(2003)\citenamefont {Wang},
  \citenamefont {Rogado}, \citenamefont {Cava},\ and\ \citenamefont
  {Ong}}]{Ong1}%
  \BibitemOpen
  \bibfield  {author} {\bibinfo {author} {\bibfnamefont {Y.}~\bibnamefont
  {Wang}}, \bibinfo {author} {\bibfnamefont {N.~S.}\ \bibnamefont {Rogado}},
  \bibinfo {author} {\bibfnamefont {R.~J.}\ \bibnamefont {Cava}}, \ and\
  \bibinfo {author} {\bibfnamefont {N.~P.}\ \bibnamefont {Ong}},\ }\bibfield
  {title} {\enquote {\bibinfo {title} {Spin entropy as the likely source of
  enhanced thermopower in na$_x$co$_2$o$_4$},}\ }\href
  {http://dx.doi.org/10.1038/nature01639} {\bibfield  {journal} {\bibinfo
  {journal} {Nature}\ }\textbf {\bibinfo {volume} {423}},\ \bibinfo {pages}
  {425} (\bibinfo {year} {2003})}\BibitemShut {NoStop}%
\bibitem [{\citenamefont {Shibauchi}\ \emph {et~al.}(2014)\citenamefont
  {Shibauchi}, \citenamefont {Carrington},\ and\ \citenamefont
  {Matsuda}}]{Matsuda14}%
  \BibitemOpen
  \bibfield  {author} {\bibinfo {author} {\bibfnamefont {T.}~\bibnamefont
  {Shibauchi}}, \bibinfo {author} {\bibfnamefont {A.}~\bibnamefont
  {Carrington}}, \ and\ \bibinfo {author} {\bibfnamefont {Y.}~\bibnamefont
  {Matsuda}},\ }\bibfield  {title} {\enquote {\bibinfo {title} {A quantum
  critical point lying beneath the superconducting dome in iron pnictides},}\
  }\href {\doibase 10.1146/annurev-conmatphys-031113-133921} {\bibfield
  {journal} {\bibinfo  {journal} {Annual Review of Condensed Matter Physics}\
  }\textbf {\bibinfo {volume} {5}},\ \bibinfo {pages} {113} (\bibinfo {year}
  {2014})}\BibitemShut {NoStop}%
\bibitem [{\citenamefont {Takagi}\ \emph {et~al.}(1992)\citenamefont {Takagi},
  \citenamefont {Batlogg}, \citenamefont {Kao}, \citenamefont {Kwo},
  \citenamefont {Cava}, \citenamefont {Krajewski},\ and\ \citenamefont
  {Peck}}]{Takagi}%
  \BibitemOpen
  \bibfield  {author} {\bibinfo {author} {\bibfnamefont {H.}~\bibnamefont
  {Takagi}}, \bibinfo {author} {\bibfnamefont {B.}~\bibnamefont {Batlogg}},
  \bibinfo {author} {\bibfnamefont {H.~L.}\ \bibnamefont {Kao}}, \bibinfo
  {author} {\bibfnamefont {J.}~\bibnamefont {Kwo}}, \bibinfo {author}
  {\bibfnamefont {R.~J.}\ \bibnamefont {Cava}}, \bibinfo {author}
  {\bibfnamefont {J.~J.}\ \bibnamefont {Krajewski}}, \ and\ \bibinfo {author}
  {\bibfnamefont {W.~F.}\ \bibnamefont {Peck}},\ }\bibfield  {title} {\enquote
  {\bibinfo {title} {Systematic evolution of temperature-dependent resistivity
  in
  ${\mathrm{la}}_{2\mathrm{\ensuremath{-}}\mathit{x}}$${\mathrm{sr}}_{\mathit{x}}$${\mathrm{cuo}}_{4}$},}\
  }\href {\doibase 10.1103/PhysRevLett.69.2975} {\bibfield  {journal} {\bibinfo
   {journal} {Phys. Rev. Lett.}\ }\textbf {\bibinfo {volume} {69}},\ \bibinfo
  {pages} {2975} (\bibinfo {year} {1992})}\BibitemShut {NoStop}%
\bibitem [{\citenamefont {Giraldo-Gallo}\ \emph {et~al.}(2018)\citenamefont
  {Giraldo-Gallo}, \citenamefont {Galvis}, \citenamefont {Stegen},
  \citenamefont {Modic}, \citenamefont {Balakirev}, \citenamefont {Betts},
  \citenamefont {Lian}, \citenamefont {Moir}, \citenamefont {Riggs},
  \citenamefont {Wu}, \citenamefont {Bollinger}, \citenamefont {He},
  \citenamefont {Bo{\v z}ovi{\'c}}, \citenamefont {Ramshaw}, \citenamefont
  {McDonald}, \citenamefont {Boebinger},\ and\ \citenamefont
  {Shekhter}}]{boebinger}%
  \BibitemOpen
  \bibfield  {author} {\bibinfo {author} {\bibfnamefont {P.}~\bibnamefont
  {Giraldo-Gallo}}, \bibinfo {author} {\bibfnamefont {J.~A.}\ \bibnamefont
  {Galvis}}, \bibinfo {author} {\bibfnamefont {Z.}~\bibnamefont {Stegen}},
  \bibinfo {author} {\bibfnamefont {K.~A.}\ \bibnamefont {Modic}}, \bibinfo
  {author} {\bibfnamefont {F.~F.}\ \bibnamefont {Balakirev}}, \bibinfo {author}
  {\bibfnamefont {J.~B.}\ \bibnamefont {Betts}}, \bibinfo {author}
  {\bibfnamefont {X.}~\bibnamefont {Lian}}, \bibinfo {author} {\bibfnamefont
  {C.}~\bibnamefont {Moir}}, \bibinfo {author} {\bibfnamefont {S.~C.}\
  \bibnamefont {Riggs}}, \bibinfo {author} {\bibfnamefont {J.}~\bibnamefont
  {Wu}}, \bibinfo {author} {\bibfnamefont {A.~T.}\ \bibnamefont {Bollinger}},
  \bibinfo {author} {\bibfnamefont {X.}~\bibnamefont {He}}, \bibinfo {author}
  {\bibfnamefont {I.}~\bibnamefont {Bo{\v z}ovi{\'c}}}, \bibinfo {author}
  {\bibfnamefont {B.~J.}\ \bibnamefont {Ramshaw}}, \bibinfo {author}
  {\bibfnamefont {R.~D.}\ \bibnamefont {McDonald}}, \bibinfo {author}
  {\bibfnamefont {G.~S.}\ \bibnamefont {Boebinger}}, \ and\ \bibinfo {author}
  {\bibfnamefont {A.}~\bibnamefont {Shekhter}},\ }\bibfield  {title} {\enquote
  {\bibinfo {title} {Scale-invariant magnetoresistance in a cuprate
  superconductor},}\ }\href {\doibase 10.1126/science.aan3178} {\bibfield
  {journal} {\bibinfo  {journal} {Science}\ }\textbf {\bibinfo {volume}
  {361}},\ \bibinfo {pages} {479} (\bibinfo {year} {2018})}\BibitemShut
  {NoStop}%
\bibitem [{\citenamefont {L\"ohneysen}\ \emph {et~al.}(2007)\citenamefont
  {L\"ohneysen}, \citenamefont {Rosch}, \citenamefont {Vojta},\ and\
  \citenamefont {W\"olfle}}]{rmpqcp}%
  \BibitemOpen
  \bibfield  {author} {\bibinfo {author} {\bibfnamefont {H.~v.}\ \bibnamefont
  {L\"ohneysen}}, \bibinfo {author} {\bibfnamefont {A.}~\bibnamefont {Rosch}},
  \bibinfo {author} {\bibfnamefont {M.}~\bibnamefont {Vojta}}, \ and\ \bibinfo
  {author} {\bibfnamefont {P.}~\bibnamefont {W\"olfle}},\ }\bibfield  {title}
  {\enquote {\bibinfo {title} {Fermi-liquid instabilities at magnetic quantum
  phase transitions},}\ }\href {\doibase 10.1103/RevModPhys.79.1015} {\bibfield
   {journal} {\bibinfo  {journal} {Rev. Mod. Phys.}\ }\textbf {\bibinfo
  {volume} {79}},\ \bibinfo {pages} {1015} (\bibinfo {year}
  {2007})}\BibitemShut {NoStop}%
\bibitem [{\citenamefont {Damascelli}\ \emph {et~al.}(2003)\citenamefont
  {Damascelli}, \citenamefont {Hussain},\ and\ \citenamefont {Shen}}]{zxs}%
  \BibitemOpen
  \bibfield  {author} {\bibinfo {author} {\bibfnamefont {A.}~\bibnamefont
  {Damascelli}}, \bibinfo {author} {\bibfnamefont {Z.}~\bibnamefont {Hussain}},
  \ and\ \bibinfo {author} {\bibfnamefont {Z.-X.}\ \bibnamefont {Shen}},\
  }\bibfield  {title} {\enquote {\bibinfo {title} {Angle-resolved photoemission
  studies of the cuprate superconductors},}\ }\href {\doibase
  10.1103/RevModPhys.75.473} {\bibfield  {journal} {\bibinfo  {journal} {Rev.
  Mod. Phys.}\ }\textbf {\bibinfo {volume} {75}},\ \bibinfo {pages} {473}
  (\bibinfo {year} {2003})}\BibitemShut {NoStop}%
\bibitem [{\citenamefont {Marel}\ \emph {et~al.}(2003)\citenamefont {Marel},
  \citenamefont {Molegraaf}, \citenamefont {Zaanen}, \citenamefont {Nussinov},
  \citenamefont {Carbone}, \citenamefont {Damascelli}, \citenamefont {Eisaki},
  \citenamefont {Greven}, \citenamefont {Kes},\ and\ \citenamefont
  {Li}}]{Marel}%
  \BibitemOpen
  \bibfield  {author} {\bibinfo {author} {\bibfnamefont {D.~v.~d.}\
  \bibnamefont {Marel}}, \bibinfo {author} {\bibfnamefont {H.~J.~A.}\
  \bibnamefont {Molegraaf}}, \bibinfo {author} {\bibfnamefont {J.}~\bibnamefont
  {Zaanen}}, \bibinfo {author} {\bibfnamefont {Z.}~\bibnamefont {Nussinov}},
  \bibinfo {author} {\bibfnamefont {F.}~\bibnamefont {Carbone}}, \bibinfo
  {author} {\bibfnamefont {A.}~\bibnamefont {Damascelli}}, \bibinfo {author}
  {\bibfnamefont {H.}~\bibnamefont {Eisaki}}, \bibinfo {author} {\bibfnamefont
  {M.}~\bibnamefont {Greven}}, \bibinfo {author} {\bibfnamefont {P.~H.}\
  \bibnamefont {Kes}}, \ and\ \bibinfo {author} {\bibfnamefont
  {M.}~\bibnamefont {Li}},\ }\bibfield  {title} {\enquote {\bibinfo {title}
  {Quantum critical behaviour in a high-tc superconductor},}\ }\href
  {https://doi.org/10.1038/nature01978} {\bibfield  {journal} {\bibinfo
  {journal} {Nature}\ }\textbf {\bibinfo {volume} {425}},\ \bibinfo {pages}
  {271} (\bibinfo {year} {2003})}\BibitemShut {NoStop}%
\bibitem [{\citenamefont {Wang}\ \emph {et~al.}(2004)\citenamefont {Wang},
  \citenamefont {Yang}, \citenamefont {Sekharan}, \citenamefont {Ding},
  \citenamefont {Engelbrecht}, \citenamefont {Dai}, \citenamefont {Wang},
  \citenamefont {Kaminski}, \citenamefont {Valla}, \citenamefont {Kidd},
  \citenamefont {Fedorov},\ and\ \citenamefont {Johnson}}]{johnson}%
  \BibitemOpen
  \bibfield  {author} {\bibinfo {author} {\bibfnamefont {S.-C.}\ \bibnamefont
  {Wang}}, \bibinfo {author} {\bibfnamefont {H.-B.}\ \bibnamefont {Yang}},
  \bibinfo {author} {\bibfnamefont {A.~K.~P.}\ \bibnamefont {Sekharan}},
  \bibinfo {author} {\bibfnamefont {H.}~\bibnamefont {Ding}}, \bibinfo {author}
  {\bibfnamefont {J.~R.}\ \bibnamefont {Engelbrecht}}, \bibinfo {author}
  {\bibfnamefont {X.}~\bibnamefont {Dai}}, \bibinfo {author} {\bibfnamefont
  {Z.}~\bibnamefont {Wang}}, \bibinfo {author} {\bibfnamefont {A.}~\bibnamefont
  {Kaminski}}, \bibinfo {author} {\bibfnamefont {T.}~\bibnamefont {Valla}},
  \bibinfo {author} {\bibfnamefont {T.}~\bibnamefont {Kidd}}, \bibinfo {author}
  {\bibfnamefont {A.~V.}\ \bibnamefont {Fedorov}}, \ and\ \bibinfo {author}
  {\bibfnamefont {P.~D.}\ \bibnamefont {Johnson}},\ }\bibfield  {title}
  {\enquote {\bibinfo {title} {Quasiparticle line shape of
  ${\mathrm{sr}}_{2}{\mathrm{ruo}}_{4}$ and its relation to anisotropic
  transport},}\ }\href {\doibase 10.1103/PhysRevLett.92.137002} {\bibfield
  {journal} {\bibinfo  {journal} {Phys. Rev. Lett.}\ }\textbf {\bibinfo
  {volume} {92}},\ \bibinfo {pages} {137002} (\bibinfo {year}
  {2004})}\BibitemShut {NoStop}%
\bibitem [{\citenamefont {Zaanen}(2004)}]{Zaanen04}%
  \BibitemOpen
  \bibfield  {author} {\bibinfo {author} {\bibfnamefont {J.}~\bibnamefont
  {Zaanen}},\ }\bibfield  {title} {\enquote {\bibinfo {title}
  {Superconductivity: Why the temperature is high},}\ }\href {\doibase
  10.1038/430512a} {\bibfield  {journal} {\bibinfo  {journal} {Nature}\
  }\textbf {\bibinfo {volume} {430}},\ \bibinfo {pages} {512} (\bibinfo {year}
  {2004})}\BibitemShut {NoStop}%
\bibitem [{\citenamefont {Chowdhury}\ \emph {et~al.}(2018)\citenamefont
  {Chowdhury}, \citenamefont {Werman}, \citenamefont {Berg},\ and\
  \citenamefont {Senthil}}]{DC2018}%
  \BibitemOpen
  \bibfield  {author} {\bibinfo {author} {\bibfnamefont {D.}~\bibnamefont
  {Chowdhury}}, \bibinfo {author} {\bibfnamefont {Y.}~\bibnamefont {Werman}},
  \bibinfo {author} {\bibfnamefont {E.}~\bibnamefont {Berg}}, \ and\ \bibinfo
  {author} {\bibfnamefont {T.}~\bibnamefont {Senthil}},\ }\bibfield  {title}
  {\enquote {\bibinfo {title} {Translationally invariant non-fermi-liquid
  metals with critical fermi surfaces: Solvable models},}\ }\href {\doibase
  10.1103/PhysRevX.8.031024} {\bibfield  {journal} {\bibinfo  {journal} {Phys.
  Rev. X}\ }\textbf {\bibinfo {volume} {8}},\ \bibinfo {pages} {031024}
  (\bibinfo {year} {2018})}\BibitemShut {NoStop}%
\bibitem [{\citenamefont {Bruin}\ \emph {et~al.}(2013)\citenamefont {Bruin},
  \citenamefont {Sakai}, \citenamefont {Perry},\ and\ \citenamefont
  {Mackenzie}}]{Bruin13}%
  \BibitemOpen
  \bibfield  {author} {\bibinfo {author} {\bibfnamefont {J.~A.~N.}\
  \bibnamefont {Bruin}}, \bibinfo {author} {\bibfnamefont {H.}~\bibnamefont
  {Sakai}}, \bibinfo {author} {\bibfnamefont {R.~S.}\ \bibnamefont {Perry}}, \
  and\ \bibinfo {author} {\bibfnamefont {A.~P.}\ \bibnamefont {Mackenzie}},\
  }\bibfield  {title} {\enquote {\bibinfo {title} {Similarity of scattering
  rates in metals showing t-linear resistivity},}\ }\href {\doibase
  10.1126/science.1227612} {\bibfield  {journal} {\bibinfo  {journal}
  {Science}\ }\textbf {\bibinfo {volume} {339}},\ \bibinfo {pages} {804}
  (\bibinfo {year} {2013})}\BibitemShut {NoStop}%
\bibitem [{\citenamefont {Legros}\ \emph {et~al.}(2018)\citenamefont {Legros},
  \citenamefont {Benhabib}, \citenamefont {Tabis}, \citenamefont
  {Lalibert{\'e}}, \citenamefont {Dion}, \citenamefont {Lizaire}, \citenamefont
  {Vignolle}, \citenamefont {Vignolles}, \citenamefont {Raffy}, \citenamefont
  {Li}, \citenamefont {Auban-Senzier}, \citenamefont {Doiron-Leyraud},
  \citenamefont {Fournier}, \citenamefont {Colson}, \citenamefont {Taillefer},\
  and\ \citenamefont {Proust}}]{Taillefer18}%
  \BibitemOpen
  \bibfield  {author} {\bibinfo {author} {\bibfnamefont {A.}~\bibnamefont
  {Legros}}, \bibinfo {author} {\bibfnamefont {S.}~\bibnamefont {Benhabib}},
  \bibinfo {author} {\bibfnamefont {W.}~\bibnamefont {Tabis}}, \bibinfo
  {author} {\bibfnamefont {F.}~\bibnamefont {Lalibert{\'e}}}, \bibinfo {author}
  {\bibfnamefont {M.}~\bibnamefont {Dion}}, \bibinfo {author} {\bibfnamefont
  {M.}~\bibnamefont {Lizaire}}, \bibinfo {author} {\bibfnamefont
  {B.}~\bibnamefont {Vignolle}}, \bibinfo {author} {\bibfnamefont
  {D.}~\bibnamefont {Vignolles}}, \bibinfo {author} {\bibfnamefont
  {H.}~\bibnamefont {Raffy}}, \bibinfo {author} {\bibfnamefont {Z.~Z.}\
  \bibnamefont {Li}}, \bibinfo {author} {\bibfnamefont {P.}~\bibnamefont
  {Auban-Senzier}}, \bibinfo {author} {\bibfnamefont {N.}~\bibnamefont
  {Doiron-Leyraud}}, \bibinfo {author} {\bibfnamefont {P.}~\bibnamefont
  {Fournier}}, \bibinfo {author} {\bibfnamefont {D.}~\bibnamefont {Colson}},
  \bibinfo {author} {\bibfnamefont {L.}~\bibnamefont {Taillefer}}, \ and\
  \bibinfo {author} {\bibfnamefont {C.}~\bibnamefont {Proust}},\ }\bibfield
  {title} {\enquote {\bibinfo {title} {Universal t-linear resistivity and
  planckian dissipation in overdoped cuprates},}\ }\href {\doibase
  10.1038/s41567-018-0334-2} {\bibfield  {journal} {\bibinfo  {journal} {Nature
  Physics}\ }\textbf {\bibinfo {volume} {14}} (\bibinfo {year} {2018}),\
  10.1038/s41567-018-0334-2}\BibitemShut {NoStop}%
\bibitem [{\citenamefont {Cao}\ \emph {et~al.}(2018{\natexlab{a}})\citenamefont
  {Cao}, \citenamefont {Fatemi}, \citenamefont {Demir}, \citenamefont {Fang},
  \citenamefont {Tomarken}, \citenamefont {Luo}, \citenamefont
  {Sanchez-Yamagishi}, \citenamefont {Watanabe}, \citenamefont {Taniguchi},
  \citenamefont {Kaxiras}, \citenamefont {Ashoori},\ and\ \citenamefont
  {Jarillo-Herrero}}]{Cao2018a}%
  \BibitemOpen
  \bibfield  {author} {\bibinfo {author} {\bibfnamefont {Y.}~\bibnamefont
  {Cao}}, \bibinfo {author} {\bibfnamefont {V.}~\bibnamefont {Fatemi}},
  \bibinfo {author} {\bibfnamefont {A.}~\bibnamefont {Demir}}, \bibinfo
  {author} {\bibfnamefont {S.}~\bibnamefont {Fang}}, \bibinfo {author}
  {\bibfnamefont {S.~L.}\ \bibnamefont {Tomarken}}, \bibinfo {author}
  {\bibfnamefont {J.~Y.}\ \bibnamefont {Luo}}, \bibinfo {author} {\bibfnamefont
  {J.~D.}\ \bibnamefont {Sanchez-Yamagishi}}, \bibinfo {author} {\bibfnamefont
  {K.}~\bibnamefont {Watanabe}}, \bibinfo {author} {\bibfnamefont
  {T.}~\bibnamefont {Taniguchi}}, \bibinfo {author} {\bibfnamefont
  {E.}~\bibnamefont {Kaxiras}}, \bibinfo {author} {\bibfnamefont {R.~C.}\
  \bibnamefont {Ashoori}}, \ and\ \bibinfo {author} {\bibfnamefont
  {P.}~\bibnamefont {Jarillo-Herrero}},\ }\bibfield  {title} {\enquote
  {\bibinfo {title} {Correlated insulator behaviour at half-filling in
  magic-angle graphene superlattices},}\ }\href
  {http://dx.doi.org/10.1038/nature26154} {\bibfield  {journal} {\bibinfo
  {journal} {Nature}\ }\textbf {\bibinfo {volume} {556}},\ \bibinfo {pages} {80
  } (\bibinfo {year} {2018}{\natexlab{a}})}\BibitemShut {NoStop}%
\bibitem [{\citenamefont {Cao}\ \emph {et~al.}(2018{\natexlab{b}})\citenamefont
  {Cao}, \citenamefont {Fatemi}, \citenamefont {Fang}, \citenamefont
  {Watanabe}, \citenamefont {Taniguchi}, \citenamefont {Kaxiras},\ and\
  \citenamefont {Jarillo-Herrero}}]{Cao2018b}%
  \BibitemOpen
  \bibfield  {author} {\bibinfo {author} {\bibfnamefont {Y.}~\bibnamefont
  {Cao}}, \bibinfo {author} {\bibfnamefont {V.}~\bibnamefont {Fatemi}},
  \bibinfo {author} {\bibfnamefont {S.}~\bibnamefont {Fang}}, \bibinfo {author}
  {\bibfnamefont {K.}~\bibnamefont {Watanabe}}, \bibinfo {author}
  {\bibfnamefont {T.}~\bibnamefont {Taniguchi}}, \bibinfo {author}
  {\bibfnamefont {E.}~\bibnamefont {Kaxiras}}, \ and\ \bibinfo {author}
  {\bibfnamefont {P.}~\bibnamefont {Jarillo-Herrero}},\ }\bibfield  {title}
  {\enquote {\bibinfo {title} {Unconventional superconductivity in magic-angle
  graphene superlattices},}\ }\href {http://dx.doi.org/10.1038/nature26160}
  {\bibfield  {journal} {\bibinfo  {journal} {Nature}\ }\textbf {\bibinfo
  {volume} {556}},\ \bibinfo {pages} {43 } (\bibinfo {year}
  {2018}{\natexlab{b}})}\BibitemShut {NoStop}%
\bibitem [{\citenamefont {Lopes~dos Santos}\ \emph {et~al.}(2007)\citenamefont
  {Lopes~dos Santos}, \citenamefont {Peres},\ and\ \citenamefont
  {Castro~Neto}}]{Castro}%
  \BibitemOpen
  \bibfield  {author} {\bibinfo {author} {\bibfnamefont {J.~M.~B.}\
  \bibnamefont {Lopes~dos Santos}}, \bibinfo {author} {\bibfnamefont
  {N.~M.~R.}\ \bibnamefont {Peres}}, \ and\ \bibinfo {author} {\bibfnamefont
  {A.~H.}\ \bibnamefont {Castro~Neto}},\ }\bibfield  {title} {\enquote
  {\bibinfo {title} {Graphene bilayer with a twist: Electronic structure},}\
  }\href {\doibase 10.1103/PhysRevLett.99.256802} {\bibfield  {journal}
  {\bibinfo  {journal} {Phys. Rev. Lett.}\ }\textbf {\bibinfo {volume} {99}},\
  \bibinfo {pages} {256802} (\bibinfo {year} {2007})}\BibitemShut {NoStop}%
\bibitem [{\citenamefont {Su\'arez~Morell}\ \emph {et~al.}(2010)\citenamefont
  {Su\'arez~Morell}, \citenamefont {Correa}, \citenamefont {Vargas},
  \citenamefont {Pacheco},\ and\ \citenamefont {Barticevic}}]{suarez}%
  \BibitemOpen
  \bibfield  {author} {\bibinfo {author} {\bibfnamefont {E.}~\bibnamefont
  {Su\'arez~Morell}}, \bibinfo {author} {\bibfnamefont {J.~D.}\ \bibnamefont
  {Correa}}, \bibinfo {author} {\bibfnamefont {P.}~\bibnamefont {Vargas}},
  \bibinfo {author} {\bibfnamefont {M.}~\bibnamefont {Pacheco}}, \ and\
  \bibinfo {author} {\bibfnamefont {Z.}~\bibnamefont {Barticevic}},\ }\bibfield
   {title} {\enquote {\bibinfo {title} {Flat bands in slightly twisted bilayer
  graphene: Tight-binding calculations},}\ }\href {\doibase
  10.1103/PhysRevB.82.121407} {\bibfield  {journal} {\bibinfo  {journal} {Phys.
  Rev. B}\ }\textbf {\bibinfo {volume} {82}},\ \bibinfo {pages} {121407}
  (\bibinfo {year} {2010})}\BibitemShut {NoStop}%
\bibitem [{\citenamefont {Bistritzer}\ and\ \citenamefont
  {MacDonald}(2011)}]{macdonald11}%
  \BibitemOpen
  \bibfield  {author} {\bibinfo {author} {\bibfnamefont {R.}~\bibnamefont
  {Bistritzer}}\ and\ \bibinfo {author} {\bibfnamefont {A.~H.}\ \bibnamefont
  {MacDonald}},\ }\bibfield  {title} {\enquote {\bibinfo {title} {Moir{\'e}
  bands in twisted double-layer graphene},}\ }\href {\doibase
  10.1073/pnas.1108174108} {\bibfield  {journal} {\bibinfo  {journal}
  {Proceedings of the National Academy of Sciences}\ }\textbf {\bibinfo
  {volume} {108}},\ \bibinfo {pages} {12233} (\bibinfo {year}
  {2011})}\BibitemShut {NoStop}%
\bibitem [{\citenamefont {Kim}\ \emph {et~al.}(2017)\citenamefont {Kim},
  \citenamefont {DaSilva}, \citenamefont {Huang}, \citenamefont {Fallahazad},
  \citenamefont {Larentis}, \citenamefont {Taniguchi}, \citenamefont
  {Watanabe}, \citenamefont {LeRoy}, \citenamefont {MacDonald},\ and\
  \citenamefont {Tutuc}}]{Tutuc}%
  \BibitemOpen
  \bibfield  {author} {\bibinfo {author} {\bibfnamefont {K.}~\bibnamefont
  {Kim}}, \bibinfo {author} {\bibfnamefont {A.}~\bibnamefont {DaSilva}},
  \bibinfo {author} {\bibfnamefont {S.}~\bibnamefont {Huang}}, \bibinfo
  {author} {\bibfnamefont {B.}~\bibnamefont {Fallahazad}}, \bibinfo {author}
  {\bibfnamefont {S.}~\bibnamefont {Larentis}}, \bibinfo {author}
  {\bibfnamefont {T.}~\bibnamefont {Taniguchi}}, \bibinfo {author}
  {\bibfnamefont {K.}~\bibnamefont {Watanabe}}, \bibinfo {author}
  {\bibfnamefont {B.~J.}\ \bibnamefont {LeRoy}}, \bibinfo {author}
  {\bibfnamefont {A.~H.}\ \bibnamefont {MacDonald}}, \ and\ \bibinfo {author}
  {\bibfnamefont {E.}~\bibnamefont {Tutuc}},\ }\bibfield  {title} {\enquote
  {\bibinfo {title} {Tunable moir{\'e} bands and strong correlations in
  small-twist-angle bilayer graphene},}\ }\href {\doibase
  10.1073/pnas.1620140114} {\bibfield  {journal} {\bibinfo  {journal}
  {Proceedings of the National Academy of Sciences}\ }\textbf {\bibinfo
  {volume} {114}},\ \bibinfo {pages} {3364} (\bibinfo {year}
  {2017})}\BibitemShut {NoStop}%
\bibitem [{\citenamefont {Cao}\ \emph {et~al.}(2016)\citenamefont {Cao},
  \citenamefont {Luo}, \citenamefont {Fatemi}, \citenamefont {Fang},
  \citenamefont {Sanchez-Yamagishi}, \citenamefont {Watanabe}, \citenamefont
  {Taniguchi}, \citenamefont {Kaxiras},\ and\ \citenamefont
  {Jarillo-Herrero}}]{Cao2016}%
  \BibitemOpen
  \bibfield  {author} {\bibinfo {author} {\bibfnamefont {Y.}~\bibnamefont
  {Cao}}, \bibinfo {author} {\bibfnamefont {J.~Y.}\ \bibnamefont {Luo}},
  \bibinfo {author} {\bibfnamefont {V.}~\bibnamefont {Fatemi}}, \bibinfo
  {author} {\bibfnamefont {S.}~\bibnamefont {Fang}}, \bibinfo {author}
  {\bibfnamefont {J.~D.}\ \bibnamefont {Sanchez-Yamagishi}}, \bibinfo {author}
  {\bibfnamefont {K.}~\bibnamefont {Watanabe}}, \bibinfo {author}
  {\bibfnamefont {T.}~\bibnamefont {Taniguchi}}, \bibinfo {author}
  {\bibfnamefont {E.}~\bibnamefont {Kaxiras}}, \ and\ \bibinfo {author}
  {\bibfnamefont {P.}~\bibnamefont {Jarillo-Herrero}},\ }\bibfield  {title}
  {\enquote {\bibinfo {title} {Superlattice-induced insulating states and
  valley-protected orbits in twisted bilayer graphene},}\ }\href {\doibase
  10.1103/PhysRevLett.117.116804} {\bibfield  {journal} {\bibinfo  {journal}
  {Phys. Rev. Lett.}\ }\textbf {\bibinfo {volume} {117}},\ \bibinfo {pages}
  {116804} (\bibinfo {year} {2016})}\BibitemShut {NoStop}%
\bibitem [{\citenamefont {{Yankowitz}}\ \emph {et~al.}(2018)\citenamefont
  {{Yankowitz}}, \citenamefont {{Chen}}, \citenamefont {{Polshyn}},
  \citenamefont {{Watanabe}}, \citenamefont {{Taniguchi}}, \citenamefont
  {{Graf}}, \citenamefont {{Young}},\ and\ \citenamefont {{Dean}}}]{dean}%
  \BibitemOpen
  \bibfield  {author} {\bibinfo {author} {\bibfnamefont {M.}~\bibnamefont
  {{Yankowitz}}}, \bibinfo {author} {\bibfnamefont {S.}~\bibnamefont {{Chen}}},
  \bibinfo {author} {\bibfnamefont {H.}~\bibnamefont {{Polshyn}}}, \bibinfo
  {author} {\bibfnamefont {K.}~\bibnamefont {{Watanabe}}}, \bibinfo {author}
  {\bibfnamefont {T.}~\bibnamefont {{Taniguchi}}}, \bibinfo {author}
  {\bibfnamefont {D.}~\bibnamefont {{Graf}}}, \bibinfo {author} {\bibfnamefont
  {A.~F.}\ \bibnamefont {{Young}}}, \ and\ \bibinfo {author} {\bibfnamefont
  {C.~R.}\ \bibnamefont {{Dean}}},\ }\bibfield  {title} {\enquote {\bibinfo
  {title} {{Tuning superconductivity in twisted bilayer graphene}},}\
  }\href@noop {} {\bibfield  {journal} {\bibinfo  {journal} {arXiv e-prints}\
  ,\ \bibinfo {eid} {arXiv:1808.07865}} (\bibinfo {year} {2018})},\ \Eprint
  {http://arxiv.org/abs/1808.07865} {arXiv:1808.07865 [cond-mat.mes-hall]}
  \BibitemShut {NoStop}%
\bibitem [{sho()}]{shoulder}%
  \BibitemOpen
  \href@noop {} {}\bibinfo {note} {Interestingly for both devices at fillings
  near $\nu=-2-\delta$ (MA1) and $\nu=2+\delta$ (MA4), we observe a broad
  shoulder around $5~$K where the resistivity drops by a substantial fraction.
  The suppression in the resistivity, before the eventual onset of SC, could be
  a signature of strong local pairing fluctuations. We will elaborate on this
  elsewhere.}\BibitemShut {Stop}%
\bibitem [{sup()}]{supp}%
  \BibitemOpen
  \href@noop {} {}\bibinfo {note} {See supplementary information for more
  details.}\BibitemShut {Stop}%
\bibitem [{\citenamefont {Chen}\ \emph {et~al.}(2008)\citenamefont {Chen},
  \citenamefont {Jang}, \citenamefont {Xiao}, \citenamefont {Ishigami},\ and\
  \citenamefont {Fuhrer}}]{monolayerT}%
  \BibitemOpen
  \bibfield  {author} {\bibinfo {author} {\bibfnamefont {J.-H.}\ \bibnamefont
  {Chen}}, \bibinfo {author} {\bibfnamefont {C.}~\bibnamefont {Jang}}, \bibinfo
  {author} {\bibfnamefont {S.}~\bibnamefont {Xiao}}, \bibinfo {author}
  {\bibfnamefont {M.}~\bibnamefont {Ishigami}}, \ and\ \bibinfo {author}
  {\bibfnamefont {M.~S.}\ \bibnamefont {Fuhrer}},\ }\bibfield  {title}
  {\enquote {\bibinfo {title} {Intrinsic and extrinsic performance limits of
  graphene devices on sio2},}\ }\href {http://dx.doi.org/10.1038/nnano.2008.58}
  {\bibfield  {journal} {\bibinfo  {journal} {Nature Nanotechnology}\ }\textbf
  {\bibinfo {volume} {3}},\ \bibinfo {pages} {206 } (\bibinfo {year}
  {2008})}\BibitemShut {NoStop}%
\bibitem [{\citenamefont {Dean}\ \emph {et~al.}(2010)\citenamefont {Dean},
  \citenamefont {Young}, \citenamefont {Meric}, \citenamefont {Lee},
  \citenamefont {Wang}, \citenamefont {Sorgenfrei}, \citenamefont {Watanabe},
  \citenamefont {Taniguchi}, \citenamefont {Kim}, \citenamefont {Shepard},\
  and\ \citenamefont {Hone}}]{mlghbn}%
  \BibitemOpen
  \bibfield  {author} {\bibinfo {author} {\bibfnamefont {C.~R.}\ \bibnamefont
  {Dean}}, \bibinfo {author} {\bibfnamefont {A.~F.}\ \bibnamefont {Young}},
  \bibinfo {author} {\bibfnamefont {I.}~\bibnamefont {Meric}}, \bibinfo
  {author} {\bibfnamefont {C.}~\bibnamefont {Lee}}, \bibinfo {author}
  {\bibfnamefont {L.}~\bibnamefont {Wang}}, \bibinfo {author} {\bibfnamefont
  {S.}~\bibnamefont {Sorgenfrei}}, \bibinfo {author} {\bibfnamefont
  {K.}~\bibnamefont {Watanabe}}, \bibinfo {author} {\bibfnamefont
  {T.}~\bibnamefont {Taniguchi}}, \bibinfo {author} {\bibfnamefont
  {P.}~\bibnamefont {Kim}}, \bibinfo {author} {\bibfnamefont {K.~L.}\
  \bibnamefont {Shepard}}, \ and\ \bibinfo {author} {\bibfnamefont
  {J.}~\bibnamefont {Hone}},\ }\bibfield  {title} {\enquote {\bibinfo {title}
  {Boron nitride substrates for high-quality graphene electronics},}\ }\href
  {https://doi.org/10.1038/nnano.2010.172} {\bibfield  {journal} {\bibinfo
  {journal} {Nature Nanotechnology}\ }\textbf {\bibinfo {volume} {5}},\
  \bibinfo {pages} {722} (\bibinfo {year} {2010})}\BibitemShut {NoStop}%
\bibitem [{\citenamefont {Das~Sarma}\ \emph {et~al.}(2011)\citenamefont
  {Das~Sarma}, \citenamefont {Adam}, \citenamefont {Hwang},\ and\ \citenamefont
  {Rossi}}]{SDSRMP}%
  \BibitemOpen
  \bibfield  {author} {\bibinfo {author} {\bibfnamefont {S.}~\bibnamefont
  {Das~Sarma}}, \bibinfo {author} {\bibfnamefont {S.}~\bibnamefont {Adam}},
  \bibinfo {author} {\bibfnamefont {E.~H.}\ \bibnamefont {Hwang}}, \ and\
  \bibinfo {author} {\bibfnamefont {E.}~\bibnamefont {Rossi}},\ }\bibfield
  {title} {\enquote {\bibinfo {title} {Electronic transport in two-dimensional
  graphene},}\ }\href {\doibase 10.1103/RevModPhys.83.407} {\bibfield
  {journal} {\bibinfo  {journal} {Rev. Mod. Phys.}\ }\textbf {\bibinfo {volume}
  {83}},\ \bibinfo {pages} {407} (\bibinfo {year} {2011})}\BibitemShut
  {NoStop}%
\bibitem [{\citenamefont {{Wu}}\ \emph {et~al.}(2018)\citenamefont {{Wu}},
  \citenamefont {{Hwang}},\ and\ \citenamefont {{Das Sarma}}}]{SDS18}%
  \BibitemOpen
  \bibfield  {author} {\bibinfo {author} {\bibfnamefont {F.}~\bibnamefont
  {{Wu}}}, \bibinfo {author} {\bibfnamefont {E.}~\bibnamefont {{Hwang}}}, \
  and\ \bibinfo {author} {\bibfnamefont {S.}~\bibnamefont {{Das Sarma}}},\
  }\bibfield  {title} {\enquote {\bibinfo {title} {{Phonon-induced giant
  linear-in-T resistivity in magic angle twisted bilayer graphene: Ordinary
  strangeness}},}\ }\href@noop {} {\bibfield  {journal} {\bibinfo  {journal}
  {arXiv e-prints}\ ,\ \bibinfo {eid} {arXiv:1811.04920}} (\bibinfo {year}
  {2018})},\ \Eprint {http://arxiv.org/abs/1811.04920} {arXiv:1811.04920
  [cond-mat.mes-hall]} \BibitemShut {NoStop}%
\bibitem [{\citenamefont {{Brown}}\ \emph {et~al.}(2018)\citenamefont
  {{Brown}}, \citenamefont {{Mitra}}, \citenamefont {{Guardado-Sanchez}},
  \citenamefont {{Nourafkan}}, \citenamefont {{Reymbaut}}, \citenamefont
  {{H{\'e}bert}}, \citenamefont {{Bergeron}}, \citenamefont {{Tremblay}},
  \citenamefont {{Kokalj}}, \citenamefont {{Huse}}, \citenamefont {{Schauss}},\
  and\ \citenamefont {{Bakr}}}]{bakr}%
  \BibitemOpen
  \bibfield  {author} {\bibinfo {author} {\bibfnamefont {P.~T.}\ \bibnamefont
  {{Brown}}}, \bibinfo {author} {\bibfnamefont {D.}~\bibnamefont {{Mitra}}},
  \bibinfo {author} {\bibfnamefont {E.}~\bibnamefont {{Guardado-Sanchez}}},
  \bibinfo {author} {\bibfnamefont {R.}~\bibnamefont {{Nourafkan}}}, \bibinfo
  {author} {\bibfnamefont {A.}~\bibnamefont {{Reymbaut}}}, \bibinfo {author}
  {\bibfnamefont {C.-D.}\ \bibnamefont {{H{\'e}bert}}}, \bibinfo {author}
  {\bibfnamefont {S.}~\bibnamefont {{Bergeron}}}, \bibinfo {author}
  {\bibfnamefont {A.~M.~S.}\ \bibnamefont {{Tremblay}}}, \bibinfo {author}
  {\bibfnamefont {J.}~\bibnamefont {{Kokalj}}}, \bibinfo {author}
  {\bibfnamefont {D.~A.}\ \bibnamefont {{Huse}}}, \bibinfo {author}
  {\bibfnamefont {P.}~\bibnamefont {{Schauss}}}, \ and\ \bibinfo {author}
  {\bibfnamefont {W.~S.}\ \bibnamefont {{Bakr}}},\ }\bibfield  {title}
  {\enquote {\bibinfo {title} {{Bad metallic transport in a cold atom
  Fermi-Hubbard system}},}\ }\href@noop {} {\bibfield  {journal} {\bibinfo
  {journal} {arXiv e-prints}\ ,\ \bibinfo {eid} {arXiv:1802.09456}} (\bibinfo
  {year} {2018})},\ \Eprint {http://arxiv.org/abs/1802.09456} {arXiv:1802.09456
  [cond-mat.quant-gas]} \BibitemShut {NoStop}%
\bibitem [{\citenamefont {{Mousatov}}\ \emph {et~al.}(2018)\citenamefont
  {{Mousatov}}, \citenamefont {{Esterlis}},\ and\ \citenamefont
  {{Hartnoll}}}]{Hartnoll18}%
  \BibitemOpen
  \bibfield  {author} {\bibinfo {author} {\bibfnamefont {C.~H.}\ \bibnamefont
  {{Mousatov}}}, \bibinfo {author} {\bibfnamefont {I.}~\bibnamefont
  {{Esterlis}}}, \ and\ \bibinfo {author} {\bibfnamefont {S.~A.}\ \bibnamefont
  {{Hartnoll}}},\ }\bibfield  {title} {\enquote {\bibinfo {title} {{Bad
  metallic transport in a modified Hubbard model}},}\ }\href@noop {} {\bibfield
   {journal} {\bibinfo  {journal} {arXiv e-prints}\ ,\ \bibinfo {eid}
  {arXiv:1803.08054}} (\bibinfo {year} {2018})},\ \Eprint
  {http://arxiv.org/abs/1803.08054} {arXiv:1803.08054 [cond-mat.str-el]}
  \BibitemShut {NoStop}%
\bibitem [{\citenamefont {Orenstein}\ \emph {et~al.}(1990)\citenamefont
  {Orenstein}, \citenamefont {Thomas}, \citenamefont {Millis}, \citenamefont
  {Cooper}, \citenamefont {Rapkine}, \citenamefont {Timusk}, \citenamefont
  {Schneemeyer},\ and\ \citenamefont {Waszczak}}]{orenstein}%
  \BibitemOpen
  \bibfield  {author} {\bibinfo {author} {\bibfnamefont {J.}~\bibnamefont
  {Orenstein}}, \bibinfo {author} {\bibfnamefont {G.~A.}\ \bibnamefont
  {Thomas}}, \bibinfo {author} {\bibfnamefont {A.~J.}\ \bibnamefont {Millis}},
  \bibinfo {author} {\bibfnamefont {S.~L.}\ \bibnamefont {Cooper}}, \bibinfo
  {author} {\bibfnamefont {D.~H.}\ \bibnamefont {Rapkine}}, \bibinfo {author}
  {\bibfnamefont {T.}~\bibnamefont {Timusk}}, \bibinfo {author} {\bibfnamefont
  {L.~F.}\ \bibnamefont {Schneemeyer}}, \ and\ \bibinfo {author} {\bibfnamefont
  {J.~V.}\ \bibnamefont {Waszczak}},\ }\bibfield  {title} {\enquote {\bibinfo
  {title} {Frequency- and temperature-dependent conductivity in
  ${\mathrm{yba}}_{2}$${\mathrm{cu}}_{3}$${\mathrm{o}}_{6+\mathit{x}}$
  crystals},}\ }\href {\doibase 10.1103/PhysRevB.42.6342} {\bibfield  {journal}
  {\bibinfo  {journal} {Phys. Rev. B}\ }\textbf {\bibinfo {volume} {42}},\
  \bibinfo {pages} {6342} (\bibinfo {year} {1990})}\BibitemShut {NoStop}%
\bibitem [{\citenamefont {Homes}\ \emph {et~al.}(2004)\citenamefont {Homes},
  \citenamefont {Dordevic}, \citenamefont {Strongin}, \citenamefont {Bonn},
  \citenamefont {Liang}, \citenamefont {Hardy}, \citenamefont {Komiya},
  \citenamefont {Ando}, \citenamefont {Yu}, \citenamefont {Kaneko},
  \citenamefont {Zhao}, \citenamefont {Greven}, \citenamefont {Basov},\ and\
  \citenamefont {Timusk}}]{Homes}%
  \BibitemOpen
  \bibfield  {author} {\bibinfo {author} {\bibfnamefont {C.~C.}\ \bibnamefont
  {Homes}}, \bibinfo {author} {\bibfnamefont {S.~V.}\ \bibnamefont {Dordevic}},
  \bibinfo {author} {\bibfnamefont {M.}~\bibnamefont {Strongin}}, \bibinfo
  {author} {\bibfnamefont {D.~A.}\ \bibnamefont {Bonn}}, \bibinfo {author}
  {\bibfnamefont {R.}~\bibnamefont {Liang}}, \bibinfo {author} {\bibfnamefont
  {W.~N.}\ \bibnamefont {Hardy}}, \bibinfo {author} {\bibfnamefont
  {S.}~\bibnamefont {Komiya}}, \bibinfo {author} {\bibfnamefont
  {Y.}~\bibnamefont {Ando}}, \bibinfo {author} {\bibfnamefont {G.}~\bibnamefont
  {Yu}}, \bibinfo {author} {\bibfnamefont {N.}~\bibnamefont {Kaneko}}, \bibinfo
  {author} {\bibfnamefont {X.}~\bibnamefont {Zhao}}, \bibinfo {author}
  {\bibfnamefont {M.}~\bibnamefont {Greven}}, \bibinfo {author} {\bibfnamefont
  {D.~N.}\ \bibnamefont {Basov}}, \ and\ \bibinfo {author} {\bibfnamefont
  {T.}~\bibnamefont {Timusk}},\ }\bibfield  {title} {\enquote {\bibinfo {title}
  {A universal scaling relation in high-temperature superconductors},}\ }\href
  {http://dx.doi.org/10.1038/nature02673} {\bibfield  {journal} {\bibinfo
  {journal} {Nature}\ }\textbf {\bibinfo {volume} {430}},\ \bibinfo {pages}
  {539} (\bibinfo {year} {2004})}\BibitemShut {NoStop}%
\bibitem [{\citenamefont {{Hazra}}\ \emph {et~al.}(2018)\citenamefont
  {{Hazra}}, \citenamefont {{Verma}},\ and\ \citenamefont
  {{Randeria}}}]{Randeria}%
  \BibitemOpen
  \bibfield  {author} {\bibinfo {author} {\bibfnamefont {T.}~\bibnamefont
  {{Hazra}}}, \bibinfo {author} {\bibfnamefont {N.}~\bibnamefont {{Verma}}}, \
  and\ \bibinfo {author} {\bibfnamefont {M.}~\bibnamefont {{Randeria}}},\
  }\bibfield  {title} {\enquote {\bibinfo {title} {{Upper bounds on the
  superfluid stiffness and superconducting \$T\_c\$: Applications to
  twisted-bilayer graphene and ultra-cold Fermi gases}},}\ }\href@noop {}
  {\bibfield  {journal} {\bibinfo  {journal} {arXiv e-prints}\ ,\ \bibinfo
  {eid} {arXiv:1811.12428}} (\bibinfo {year} {2018})},\ \Eprint
  {http://arxiv.org/abs/1811.12428} {arXiv:1811.12428 [cond-mat.supr-con]}
  \BibitemShut {NoStop}%
\end{thebibliography}%

\begin{widetext}
\renewcommand{\thefigure}{S\arabic{figure}}
\renewcommand{\figurename}{Supplemental Figure}
\setcounter{figure}{0}
\section*{Supplementary materials}
\subsection{Methods}
{\bf Sample preparation:} The six devices reported consist of two graphene and two hexagonal boron nitride layers each. Both materials were exfoliated on SiO$_2$/Si chips and the desired high-quality flakes were selected using both optical and atomic force microscopy. The fabrication of the heterostructures is based on a modified polymer-based dry pick-up technique. A poly(bisphenol A carbonate)(PC)/polydimethylsiloxane(PMMA) is placed on a glass slide and attached to a micro-positioning stage. We first pick a hexagonal boron nitride flake at around 100$^0$ C. Apart from encapsulating the graphene device to protect it from contamination, it is also used to tear the graphene flake apart thanks to the mutual attraction due to van der Waals forces. The separated piece of graphene is then rotated and stacked at room temperature. Finally, the 3-layer heterostructure is transferred onto another hexagonal boron nitride flake placed on top of a metallic gate at a temperature of 160$^0$ C. The definitive device geometry is obtained using electron beam lithography and reactive ion etching, whereas electrical connections are done with thermally evaporated Cr/Au edge-contacted guides. In addition to a Pd/Au back gate, device MA3 also incorporates a metallic top gate for a finer tuning of the carrier density of the sample.\\

{\bf Measurements:} Electronic transport measurements are carried out in a dilution refrigerator equipped with a superconducting magnet. As described in previous work  \cite{Cao2018a}, a base temperature of 70mK is achieved and low-frequency lock-in techniques are used to obtain the data. Both the current flowing through the sample and the four-probe voltages are amplified during measurements. 
In order to determine the twist angle for each device, Wannier diagrams are obtained using transport measurements and the Landau levels that appear are fitted  \cite{Cao2018a}. This procedure gives an uncertainty of about 0.02 degrees.\\

\subsection{Supplementary Text}

{\bf Theoretical estimate of the resistivity slope:} Let us now estimate the slope of the resistivity, $A$, assuming that the $T-$linear resistivity arises as a result of electron-electron interactions. For MABLG, there are three relevant energy scales --- the band width ($W$), the super-lattice gap to other higher-energy bands ($E_0$), and the interaction strength ($U$). Experimentally, it is known that $E_0$ is large \cite{Cao2016,Cao2018a} and so we consider the limit $E_0 \gg W, U$ (we can formally consider the limit $E_0\rightarrow\infty$). However, as emphasized earlier, $U$ is comparable to $W$.

We now express the resistivity as $\rho = \rho_0 + AT$,  where $\rho_0$ is affected by disorder in the sample, but where we assume that $A$ is largely unaffected by disorder. In order to estimate $A$, it is clear from dimensional analysis that $[A]=[$quantum of resistance$]/[$energy$]$. In the high-temperature regime, taking into account the degeneracy from the spin and valley degrees of freedom ($g=4$), we may write
\beq
A = \frac{h}{4e^2} \frac{1}{W} F\bigg(\frac{W}{U}, \nu,...\bigg), 
\eeq
where $F(...)$ is a function of $W/U$, the filling $\nu$ and, in principle, other quantities including the twist-angle (which affects $W$). 

However, when $W$ is comparable to $U$ and $F(\nu,...)$ does not depend strongly on $\nu$, we expect $F(\nu)\sim O(1)$ number. This leads to an estimated slope  of $A\approx h/(4e^2W)$. For a realistic estimate of $W \sim 10$ meV, we get $A\approx 60~\Omega/$K, which is similar to the measured values  up to $O(1)$ factors. Interestingly, in the strange metal regime of optimally doped cuprates, a similar estimate with $g=2$ leads to an expected slope of $A\sim h/(2e^2J)$, where $J$ represents the strength of the antiferromagnetic exchange interaction, which is close to the measured value.\\

{\bf Experimental measurement of slope of resistivity:} We have evaluated the slope, $A=d\rho(T)/dT$, from a linear fit to the resistivity data for $T>8~$K. 
In Fig.\ref{slopes}, we show a few representative fits for fillings near $\nu=-2$. 
\begin{figure*}[h]
\begin{center}
\includegraphics[scale=0.30]{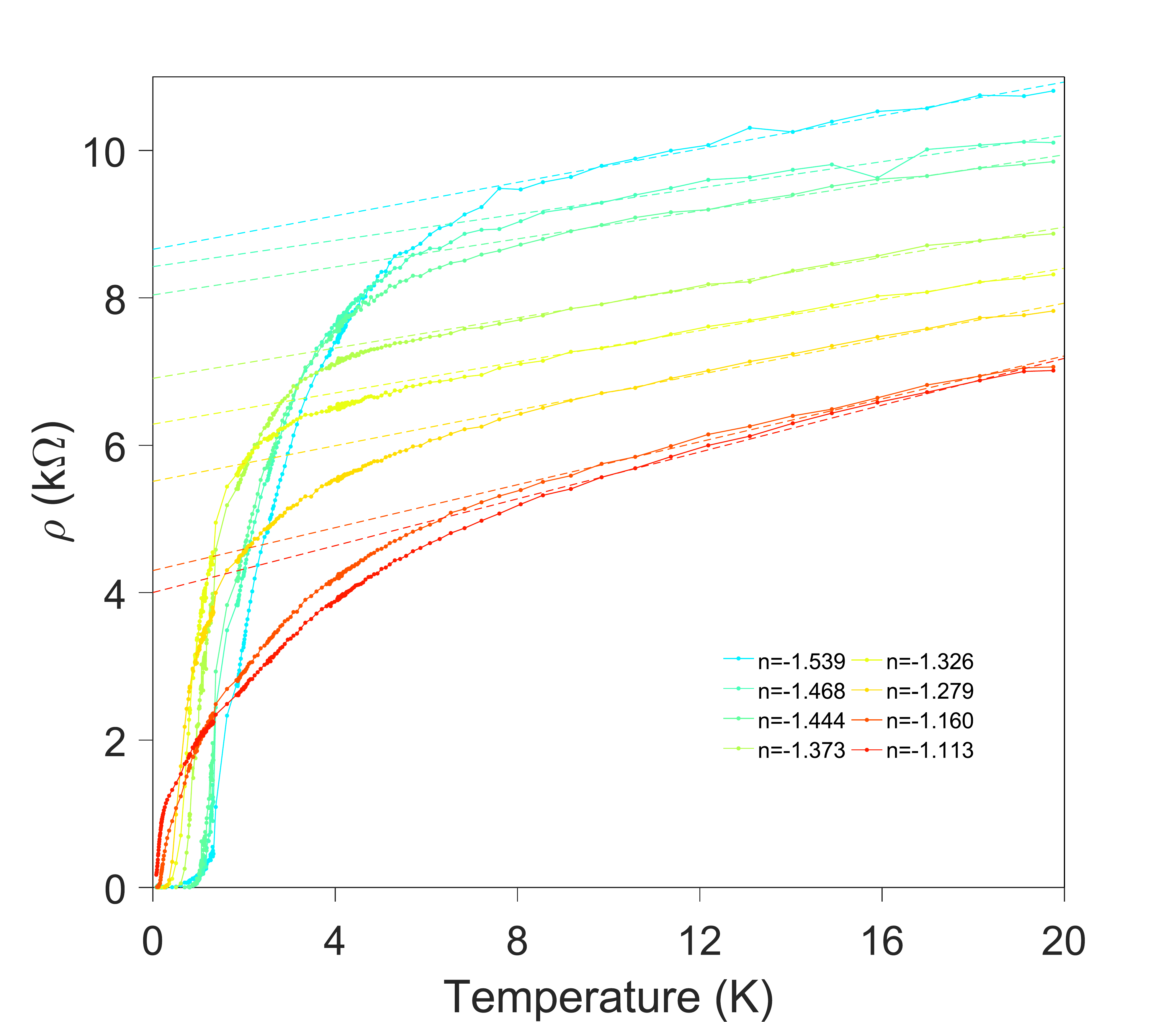}
\end{center}
\caption{{\bf Linear fits for device MA2.} 
A least squares linear fit for the resistivity, $\rho(T)$, for $T>8~$K for a few representative fillings near $\nu=-2$. The gate controlled carrier densities are denoted in the legend in units of $10^{12}$ cm$^{-2}$. }
\label{slopes}
\end{figure*}
In order to compute the coefficient $C$ from the measured slope in units of $\Omega/$K, we need the following combination of units in order to obtain a dimensionless number,
\beq
1\frac{\Omega}{\tn{K}}\times\frac{1}{m_e}\times e^2 \times \frac{\hbar}{k_B} \times 10^{12} ~cm^{-2} \approx 2.15\times 10^{-3}.
\eeq
\\ 

{\bf Phase stiffness:} Based on the existing quantum oscillation data \cite{Cao2018b} at low temperatures, we may try to address several questions related to the superconductor that emerges out of the metallic state. In particular, since we know the effective carrier concentration, $n_c$ (from the frequencies measured in SdH oscillations) and the effective mass, $m^*$, we may immediately try to infer the behavior of the `phase stiffness' (related to the zero temperature penetration depth by $D_S\propto\lambda_L^{-2}(0)$), $D_S\sim n_c/m^*$, of the superconducting state. The phase stiffness will determine how susceptible the superconductor is to thermal phase fluctuations. Naively, we expect that in superconductors that arise from doping a Mott insulator, the phase stiffness vanishes upon approaching the Mott insulator (since the number of carriers that may participate in pairing vanish in that limit). The onset of long-range phase coherence and $T_c$ is then determined by the phase stiffness, which determines the scale at which the phase of the order parameter locks in over long distances. Indeed, in the hole-doped cuprates, the phase stiffness on the underdoped side of the phase diagram exhibits such behavior \cite{orenstein}. 
However, in MABLG, for $\nu=-2-\delta$, the effective mass vanishes as $\delta\rightarrow0$ (see Fig. \ref{Tf}), as does the carrier density. Therefore we are interested in the behavior of the ratio, $D_S$, especially in the vicinity of the correlated insulator.

Before attempting to infer the behavior of $D_S$, let us review a more general principle. While there is no optical conductivity data available on MABLG at present, it is useful to place these measurements in the context of $D_S$. When a metal undergoes a SC transition, the low energy optical weight for frequencies $\omega<2\Delta$ has to be rearranged dramatically ($\Delta$ is the SC gap). It is natural for all of this weight to go into the delta-function at zero-frequencies (we are assuming that there is no appreciable spectral weight transfer to $\omega\gg\Delta$), which precisely measures the phase stiffness, i.e. the `missing-area' is given by,
\beq
D_S \approx \frac{2\Gamma}{\pi}\int_0^{2\Delta/\Gamma}~\sigma(z)~dz,
\eeq
where $\Gamma$ is the appropriate scattering rate which includes the effects of disorder, and $z\equiv \omega/\Gamma$. When $\Delta/\Gamma\gg 1$, the above reduces to $D_S = D_N$, where $D_N$
is the conductivity spectral weight (or the `Drude-weight') in the normal state. On the other hand, when $\Delta/\Gamma\ll 1$, the above relation can be simplified to yield the relation $D_S \approx \sigma(0)\Delta$, with $\sigma(0)$ the extrapolated $T\rightarrow0$ value of the conductivity from above $T>T_c$. A number of superconductors satisfy the above relationship \cite{Homes} . In this regime, $D_S$ can be substantially smaller than $D_N$, meaning that only a tiny fraction of the normal state spectral weight, compared to the width of the conductivity ($\omega\leq 2\Delta \ll 2\Gamma$), condenses into the delta-function below $T_c$. Since we do not know $\Delta$ in the experiments on MABLG, it may be reasonable to use $T_c/\Gamma$ as a proxy instead.  

It is more useful to express the expected phase stiffness (i.e. $n_c/m^*$) in terms of an energy scale, which in two-dimensions is simply the Fermi-temperature, $T_F=(2\pi \hbar^2 n_c)/(k_Bgm^*) $. In this expression, $g$ is the degeneracy factor; based on the low-temperature SdH oscillations \cite{Cao2018b} for $\nu=-2-\delta$ and $\nu=-2+\delta$ we use $g=2$ and $g=4$, respectively. Note that the combination
\beq
\frac{\pi \hbar^2}{k_B}\frac{10^{12}~\tn{cm}^{-2}}{m_e} = 27.78~\tn{K}.
\eeq 
We plot the Fermi temperature in Fig.\ref{Tf}. For fillings $\nu=-2-\delta$, we find that $T_F$ is roughly a constant independent of $n$. A naive extrapolation all the way to the correlated insulating filling ($n\approx -1.3\times 10^{12}~\tn{cm}^{-2}$) would seem to suggest that the superfluid stiffness remains finite, even as one approaches the insulator. This implies that superconductivity is lost not due to loss of phase-coherence, but due to destruction of the pairing amplitude It is however worth pointing out that it is, in principle, possible for $D_S$ to collapse dramatically as one approaches the insulator at $\nu=-2$. We also note that Ref.~\cite{Randeria} has recently proposed an upper bound on the phase stiffness in superconducting MABLG, which is approximately a tenth of our estimated $T_F$.

\begin{figure*}[h]
\begin{center}
\includegraphics[scale=0.3]{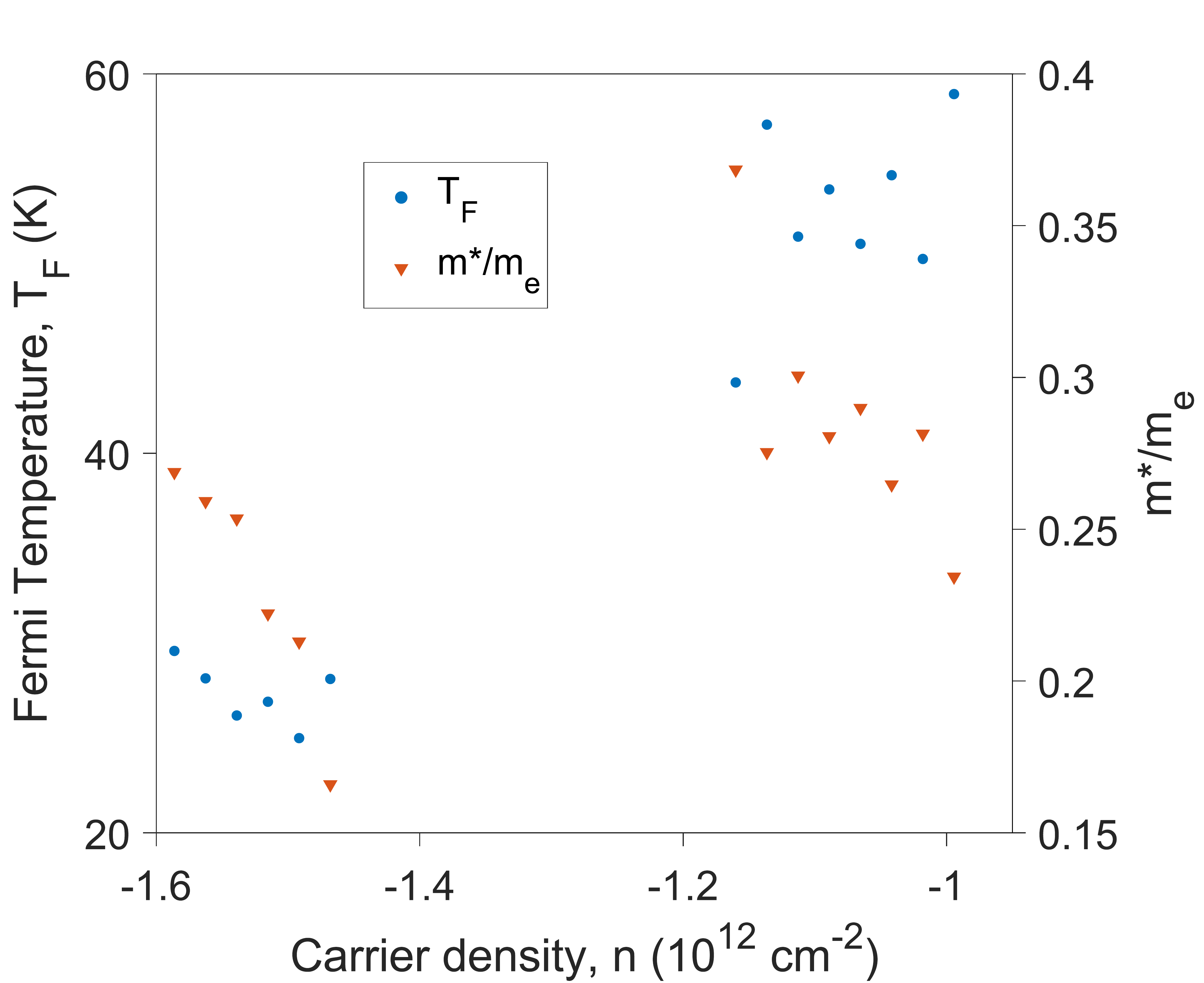}
\end{center}
\caption{{\bf Fermi temperature versus carrier density.} Left: The calculated $T_F$ (blue circles), with an assumed degeneracy of $g=2$ for $\nu=-2-\delta$ and $g=4$ for $\nu=-2+\delta$, respectively. The correlated insulator at $\nu=-2$ is located in the vicinity of $n\approx -1.3\times 10^{12} ~\tn{cm}^{-2}$. Right: The effective mass, $m^*/m_e$, obtained from SdH oscillations (orange inverted triangles) \cite{Cao2018b}. }
\label{Tf}
\end{figure*}

\end{widetext}

\end{document}